%% file: main.tex
\documentclass[
    twocolumn,
    superscriptaddress,
    amsmath,amstex,amssymb,
    aps, prx,
    floatfix,
    longbibliography,
    reprint]{revtex4-2}

\usepackage{graphicx}
\usepackage{dcolumn}%
\usepackage{bm}
\usepackage{lipsum}
\usepackage{gensymb}
\usepackage{physics}
\usepackage{float}

\newcommand{\MyTitle}{Programmable Superradiance in an Interacting Qubit Array}

\usepackage{hyperref}
\hypersetup{
  colorlinks   = true, 
  urlcolor     = blue, 
  linkcolor    = red, 
  citecolor   = red 
}

\begin{document}

\title{\MyTitle}

\author{Botao Du}
 \altaffiliation{These authors contributed equally to this work}
\author{Qihao Guo}%
\altaffiliation{These authors contributed equally to this work}
\author{Ruichao Ma}%
 \email{maruichao@purdue.edu}
\affiliation{%
Department of Physics and Astronomy, Purdue University, West Lafayette, IN 47907, USA }%

\date{\today}

\begin{abstract}

When multiple quantum emitters couple to a common electromagnetic environment, interference in their collective radiative dynamics gives rise to superradiance and subradiance. In regimes where coherent interactions and collective dissipation compete, the microscopic many-body dynamics and quantum correlations among the emitters that underlie superradiance and subradiance are theoretically challenging and remain experimentally elusive, even though collective emission has been observed in many physical systems. Here, we realize a superconducting qubit array coupled to a common microwave waveguide that mediates collective dissipation, with simultaneous access to coherent interactions and microscopic measurements of many-body dynamics. Engineered qubit-waveguide couplings with tunable amplitude and phase enable control of collective interference and the resulting super- and subradiant states. Leveraging site-resolved control and readout, we directly observe the microscopic decay dynamics of multi-qubit states across different excitation manifolds and track the evolution of populations and tunable quantum correlations. We reveal collective decay in regimes beyond the ideal Dicke model, where strong qubit-qubit interactions stabilize superradiance and subradiance against local dephasing and reshape decay pathways through spatially and spectrally structured many-body eigenstates. Our results establish a flexible platform for exploring collective phenomena in many-body quantum optics and driven-dissipative approaches to robust quantum information processing.

\end{abstract}

\maketitle

\let\oldaddcontentsline\addcontentsline
\renewcommand{\addcontentsline}[3]{}


Collective radiative decay arises from quantum interference in light-matter interactions. When multiple emitters couple coherently to a common electromagnetic environment, spontaneous emission is no longer an independent single-emitter process. Instead, interference between emission amplitudes causes them to add constructively or cancel destructively, producing enhanced or suppressed collective radiative rates known as super- and subradiance. In the Dicke model, an ensemble of identical emitters couples symmetrically to shared electromagnetic modes, giving rise to a collectively enhanced radiative decay known as Dicke superradiance \cite{Dicke1954-ae, Gross1982-pb}. This dynamics can be understood as cascaded decay through symmetric collective states, during which the emitters develop phase coherence and synchronized dipole moments. Collective coupling thus provides a paradigmatic example of how dissipation through a shared environment can generate correlated many-body dynamics, rather than acting solely as a source of decoherence \cite{Sheremet2023-ae}.

Since its original introduction, superradiance has been observed across a wide range of physical systems, including atomic gases \cite{PhysRevLett.30.309},
trapped ions \cite{PhysRevLett.76.2049}, molecules \cite{lange2024superradiant}, and solid-state nanocrystal superlattices \cite{raino2018superfluorescence}. 
In many experimental realizations, however, the idealized conditions of the Dicke model are not satisfied, leading to qualitatively different behavior and necessitating a broader theoretical framework \cite{Kirton2019-po}. Real emitter arrays generally lack perfect permutation symmetry: emitters are separated in space and arranged in structured or disordered geometries, which modify their radiative coupling \cite{Masson2022-jm, Douglas2026-jm};
their transition frequencies may be inhomogeneous; and direct coherent interactions can reshape the collective decay dynamics \cite{Mendonca2025-dg}. Dephasing further alters collective emission and can either suppress or enhance superradiance \cite{Kirton2017-ww},
while non-radiative processes compete with radiative decay. In addition, emitters often possess internal structure beyond the two-level approximation \cite{Zanner2022-ks}. In this broader regime, collective decay becomes a dynamical many-body problem in which geometry, disorder, interactions, and dissipation jointly determine whether cooperative enhancement persists and how it develops in time.

Engineered quantum systems provide a controlled setting to investigate collective dynamics in quantum emitter arrays beyond the Dicke model, particularly in waveguide and cavity quantum electrodynamics (QED) architectures with precisely tunable emitter-environment couplings \cite{Chang2018-kp, Gu2017-iu}.
Atomic arrays with tunable spacing demonstrate the role of geometry in radiative coupling \cite{Yan2023-as, Douglas2026-jm}, while nanophotonic platforms enable cascaded superradiance in chiral waveguides \cite{Liedl2024-mj} and selective collective emission via structured photonic environments \cite{Zhou2025-vs}. In superconducting artificial atoms \cite{Carusotto2020-ct, Blais2021-bn}, 
Dicke superradiance has been demonstrated for pairs of qubits \cite{Mlynek2014-ez}, 
while multi-qubit experiments have explored the manipulation of subradiant states \cite{Zanner2022-ks}, bandgap engineering in waveguides \cite{Brehm2021-eo}, and tunable collective interactions \cite{Wang2020-ag}.

Despite these advances, most experiments probe collective decay from measurements of the emitted radiation, leaving the microscopic dynamics of the emitters during the decay largely inaccessible. In particular, how correlations and entanglement emerge and evolve during collective decay \cite{Zhang2025-nf} are fundamental questions that challenge theoretical modeling and remain largely unexplored experimentally. Furthermore, the role of strong coherent interactions and their interplay with collective dissipation remains poorly understood, despite their central importance in realistic emitter arrays \cite{Kersten2026-cz}. Directly resolving the microscopic dynamics of the emitters, including the emergence of correlations and interaction-driven evolution, would provide new insight into collective behavior in interacting open quantum systems.

Here, we realize a waveguide-coupled superconducting qubit array with simultaneous access to collective dissipation, coherent interactions, and microscopic emitter dynamics. Using parametric modulation, we engineer qubit-waveguide couplings with tunable amplitude, frequency, and phase, thereby controlling collective interference and the resulting super- and subradiant states. With site-resolved control and readout, we prepare multi-qubit states across different excitation manifolds and directly probe their decay dynamics by measuring emitter populations and correlations throughout the collective emission process. In our system, strong coherent interactions between the qubits fundamentally modify the collective decay dynamics: the dissipation proceeds via cascaded decay of spectrally and spatially structured many-body eigenstates, necessitating a multimode description that goes beyond the ideal Dicke model of fully symmetric collective coupling. Furthermore, the strong interactions protect superradiant and subradiant dynamics that would otherwise be suppressed by local dephasing.
These results establish a tunable platform for investigating collective dynamics in the regime of many-body quantum optics \cite{Chang2014-yo}.

\section{Engineering tunable qubit-waveguide coupling}

\begin{figure}[!tbp]
    \includegraphics[width=1.0\columnwidth]{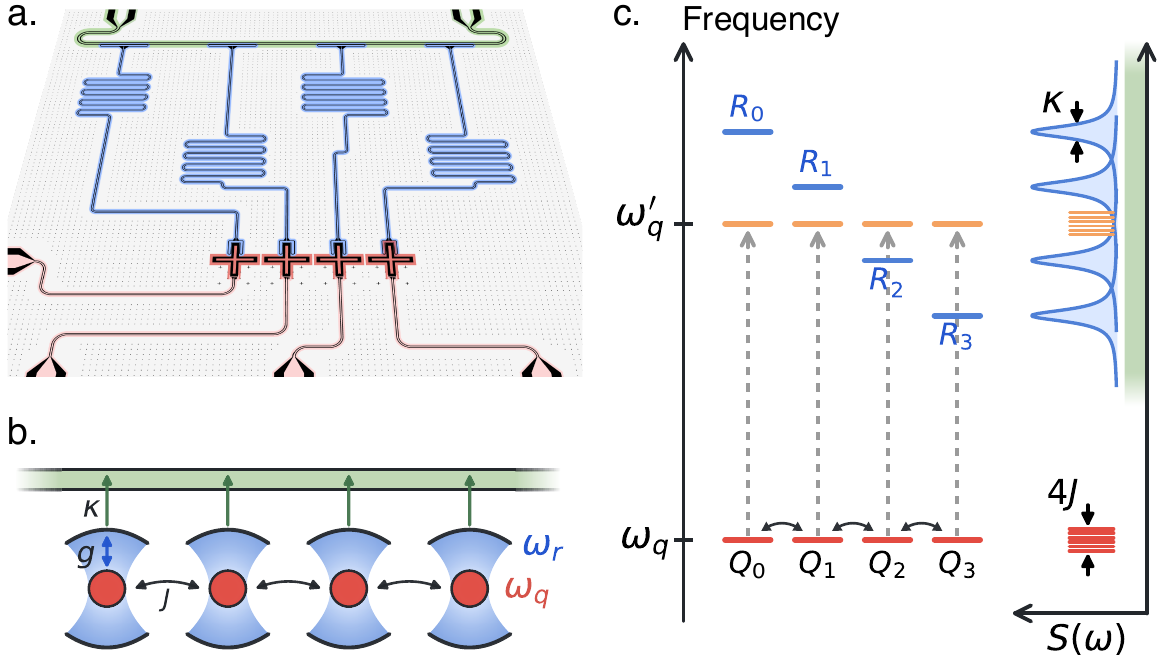}
    \caption[]{
    Waveguide-coupled qubit array and engineered dissipation.
    (a) Device layout of the superconducting circuit, comprising an array of four frequency-tunable transmon qubits (red) with local flux bias (pink), each coupled to a resonator (blue), which in turn couples to a common waveguide (green).
    (b) Effective waveguide QED model of the device.
    (c) Energy-level diagram. The qubits (red) are initially far detuned from the resonators (blue); parametric flux modulation generates qubit sidebands (orange) near the resonator frequencies, enabling enhanced Purcell decay through the resonator linewidth into the broadband waveguide, as illustrated by the spectral density $S(\omega)$. Coherent qubit–qubit interactions produce energy dispersion of the many-body eigenstates. Energies and linewidths are not to scale. 
    }
    \label{fig:1} 
\end{figure}

The experiments are performed in a superconducting circuit device as described in our previous work \cite{Du2024probing, Du2025-iw} and shown in Fig.\,\ref{fig:1}a. The effective waveguide QED setup consists of four transmon qubits, each coupled to an individual microwave cavity, which in turn are coupled to a common one-dimensional waveguide, as illustrated in Fig.\,\ref{fig:1}b.  
Each transmon qubit has a tunable frequency $\omega_q$ in the range of $2\pi\times (3.9-5.7)$\,GHz controlled via local flux bias lines. The cavities are coplanar waveguide resonators that capacitively couple to the qubits, with a qubit-resonator interaction $g(\sigma^+_i b_i + \sigma^-_i b^{\dagger}_i)$ where $g\approx 2\pi\times 65$\,MHz. Here $\sigma^+_i$ ($\sigma^-_i$) is the creation (annihilation) operator for qubit $i$ ($i=0,1,2,3$), and $b^\dagger_i$ ($b_i$) is the creation (annihilation) operator for resonator $i$. 
The resonators have linewidths $\kappa_r \approx 2\pi \times 1.5$\,MHz from their coupling to the broadband microwave transmission line with impedance-matched ports. The four resonators have distinct frequencies, $\omega_r^{0\dots3} \approx 2\pi\times(6.32, 6.28, 6.23, 6.18)$\,GHz, arranged monotonically along the qubit array.
The resonators couple to the waveguide at locations separated by a nearest-neighbor distance $d$, with $d \approx 0.03\,\lambda$ at the resonator frequencies, so position-dependent emission phases between qubits are small; here $\lambda$ is the wavelength of the emission in the waveguide.
These same resonators are also used for frequency-multiplexed dispersive readout of the qubits via the transmission line \cite{Blais2021-bn}.
In addition, the transmon qubits have direct capacitive coupling between nearest-neighbor pairs that corresponds to an exchange interaction $J(\sigma^+_i\sigma^-_{i+1}+\sigma^-_i\sigma^+_{i+1})$ with $J\approx 2\pi\times 6.0$\,MHz. 
In this work, higher levels of the transmon beyond the ground state and first excited state are not populated due to the large qubit anharmonicity $U\approx 2\pi\times -250\,\text{MHz}$. 

We start with all qubits tuned to $\omega_\text{q} \approx 2\pi\times 4.6$\,GHz, far-detuned from the resonators by $\Delta = \omega_\text{r}-\omega_\text{q} \approx 2\pi\times 1.6$\,GHz. In this dispersive limit ($\Delta \gg g$), the decay of a single qubit excitation via the resonator into the waveguide is highly suppressed, with a rate given by the Purcell limit $\kappa_\text{r} g^2/\Delta^2 \approx 2\pi \times 2.5$\,kHz. 
To engineer a tunable qubit-waveguide coupling, we modulate the frequency of each qubit by applying an AC signal through the flux bias lines. 
When the flux modulation frequency $\omega_\text{mod}$ is tuned near the resonator-qubit detuning $\Delta$, a modulation-induced sideband of the qubit spectrum at $(\omega_\text{q}+\omega_\text{mod})$ overlaps with the Lorentzian spectrum of the resonator, resulting in enhanced Purcell decay. In this engineered Purcell process, the qubit emits into the waveguide at energy $\omega_\text{q}'= (\omega_\text{q}+\omega_\text{mod})$ with a rate $\gamma_\text{wg} = (\frac{g_\text{eff}}{\Delta_\text{eff}})^2\kappa_{r}$. Here $g_\text{eff}$ is the effective modulation-induced qubit-resonator coupling rate and $\Delta_\text{eff} = \omega_{r} - (\omega_q+\omega_\text{mod})$ is the effective detuning between the resonator and relevant qubit sideband \cite{Du2025-iw}.
Hence, the parametric flux modulation enables dynamical control over the individual qubit's decay into the waveguide: the modulation frequency determines the emitted microwave energy $\omega_\text{q}'$, while the modulation amplitude determines the emission rate $\gamma_\text{wg}$. Furthermore, the relative phase of the microwave fields emitted by two qubits into the waveguide is determined by the relative phase of the applied flux modulations on the two qubits. The relevant energy levels involved in the engineered decay are illustrated in Fig.\,\ref{fig:1}c.

In the experiments that follow, we use weak, off-resonant modulation ($g_\text{eff}\ll g$, $\kappa_r \ll \Delta_\text{eff}$) to achieve qubit-waveguide coupling of typical strength $\gamma_\text{wg}\sim 2\pi\times 15$\,kHz. This engineered waveguide decay rate exceeds intrinsic qubit relaxation by a factor of $\gamma_\text{wg}/\gamma_\text{int} \sim 3$. In the absence of flux modulation, we measure single-qubit relaxation rates of $\gamma_\text{int} = 1/T_1 \approx 2\pi\times 5$\,kHz, which includes contributions from intrinsic Purcell-limited decay and non-radiative relaxation. See Supplementary Information (SI) for details on device parameters and intrinsic qubit decoherence (SI Sec.\,\ref{sec:SI_device-param}), and characterization of the modulation-induced waveguide coupling (SI Sec.\,\ref{sec:SI_eng-decay}).

\section{Collective decay of single-excitation eigenstates}

\begin{figure}[!tbp]
    \includegraphics[width=1\columnwidth]{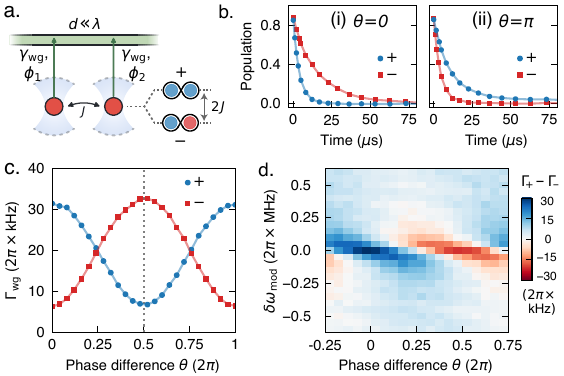}
    \caption[]{Programmable collective decay in two interacting qubits. 
    (a) Schematic of two resonant qubits \(Q_1\) and \(Q_2\) with coherent coupling \(J\), forming symmetric and antisymmetric single-excitation eigenstates \(\ket{\pm}\). Engineered waveguide coupling with equal amplitude \(\gamma_{\mathrm{wg}}\) and relative phase \(\theta=\phi_1-\phi_2\) enables interference of the emitted fields. 
    (b) Measured population decay of \(\ket{\pm}\) for (i) \(\theta=0\) and (ii) \(\theta=\pi\), showing reversal of superradiant and subradiant behavior.
    (c) Extracted engineered decay rates \(\Gamma_{\mathrm{wg}}^{\pm}\) as a function of \(\theta\), after subtracting intrinsic qubit decay, demonstrating continuous control of collective emission. 
    (d) Difference in effective decay rates \(\Gamma_{\mathrm{wg}}^{+}-\Gamma_{\mathrm{wg}}^{-}\) as a function of modulation detuning \(\delta\omega_{\mathrm{mod}}\) and phase \(\theta\). Finite detuning suppresses collective effects due to spectral distinguishability of the emitted fields.}
    \label{fig:2} 
\end{figure}

We first demonstrate tunable control of waveguide-mediated collective decay in a pair of interacting qubits by measuring the decay dynamics of single-excitation eigenstates under engineered qubit-waveguide coupling. 
We consider two neighboring qubits $Q_1$ and $Q_2$ in resonance, while the remaining qubits are far detuned and effectively inactive. In the single-excitation manifold, the eigenstates are the symmetric and antisymmetric superpositions $|\pm\rangle$ with eigenfrequencies $\omega_q \pm J$ (Fig.~\ref{fig:2}a). We prepare $|+\rangle$ ($|-\rangle$) by initially detuning the qubits, applying a resonant $\pi$-pulse to the higher- (lower-) frequency qubit, and then adiabatically reducing the qubit detuning to zero.
We then turn on the waveguide coupling by applying flux modulations to both qubits at a common modulation frequency such that the sideband-shifted qubit frequency $\omega_q'=\omega_q+\omega_\text{mod}$ lies midway between the resonator frequencies $\omega_r^1$ and $\omega_r^2$. The modulation amplitudes are chosen to give equal waveguide decay rates $\gamma_\text{wg}\approx 2\pi\times 15\,\mathrm{kHz}$ for the two qubits, while the modulation phases $\phi_1$ and $\phi_2$ set the relative coupling phase. After a variable evolution time, the modulation is turned off, and the remaining qubit population is measured. See SI Sec.\,\ref{sec:SI_expt-seq} for details on experimental sequences, state preparation, and readout.

When the waveguide couplings have uniform phases ($\theta=\phi_1-\phi_2=0$), the symmetric state $|+\rangle$ emits into the waveguide with the same phase from both qubits, leading to constructive interference and an enhanced engineered decay rate $\Gamma_\text{wg}^+ \approx 2\gamma_\text{wg}$, exceeding the single-qubit decay rate $\gamma_\text{wg}$. In contrast, the antisymmetric state $|-\rangle$ emits with opposite phase, resulting in destructive interference and a strongly suppressed engineered decay rate $\Gamma_\text{wg}^- \approx 0$.
Fig.~\ref{fig:2}b(i) shows the corresponding exponential population decay of the two states, demonstrating the contrast between superradiant and subradiant dynamics.
We define the total excitation number as \(N=\sum_i n_i\), where \(n_i=(1+\sigma_i^z)/2\) is the on-site excitation-number operator, with expectation value \(\langle N\rangle=\sum_i p_i\) and \(p_i=\langle n_i\rangle\).
When the modulations are changed to have alternating phases $\theta=\pi$, the interference condition is reversed, such that the symmetric state becomes subradiant while the antisymmetric state exhibits enhanced decay, as shown in Fig.~\ref{fig:2}b(ii).
By continuously tuning $\theta$, we programmably control the relative decay rates of the single-excitation eigenstates and realize tunable superradiant and subradiant states. Fig.~\ref{fig:2}c shows the measured engineered waveguide decay rates as a function of $\theta$, after subtracting the intrinsic qubit decay measured independently in the absence of flux modulation.

In our experiments, the coherent interaction $J$ is essential for observing collective decay by suppressing the effects of local dephasing. The measured single-qubit dephasing rate $\gamma_\phi = 1/T_2^* \approx 2\pi \times 75$\,kHz, limited by flux noise from room-temperature bias electronics, exceeds the engineered waveguide decay rates. In the absence of qubit-qubit coupling, such dephasing would randomize the relative phase between qubits and wash out the interference responsible for collective decay.
With a large coherent interaction $J \gg \gamma_\phi, \gamma_\text{wg}$, the energy splitting of $2J$ between the symmetric and antisymmetric eigenstates prevents their mixing under local dephasing. When the qubits are tuned to degeneracy, the effective flux-noise-induced linewidth of the eigenstates is suppressed to $\Gamma_\phi^\text{eff} \sim \gamma_\phi^2/(4J) \approx 2\pi\times 0.2\,\text{kHz}\ll \gamma_\text{wg}$, allowing engineered collective decay to dominate the observed dynamics. 
Previously, strong dipolar interactions have been used to spectrally resolve super- and subradiant states in molecular pairs, but in a regime where dephasing is small compared to decay \cite{Trebbia2022-vz, lange2024superradiant}.
Additional data and analysis of the role of dephasing are provided in SI Sec.\,\ref{sec:SI_extradata-qubits-detuned}.

The observed finite decay rate of the subradiant state arises from several deviations from the ideal Dicke case. Because the qubit-waveguide coupling is mediated by off-resonant, narrow-bandwidth resonators, the engineered decay rate is frequency dependent and cannot be perfectly matched for eigenstates with different energies. In addition, the resonators couple to the waveguide with a finite spatial separation $d$, leading to a position-dependent emission phase $2\pi d/\lambda$ and a corresponding reduction of the interference contrast by a factor $\cos(2\pi d/\lambda)\approx 0.985$. Theoretical modeling of the experiments, including estimation of the single-excitation decay rates, is provided in SI Sec.\,\ref{sec:SI_modeling}.

\begin{figure}[!tbp]
\includegraphics[width=0.9\columnwidth]{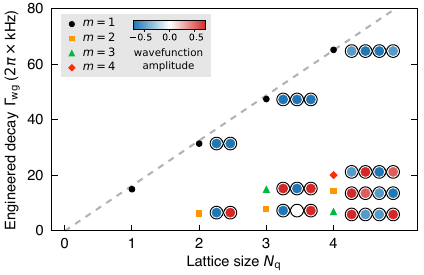}
    \caption[]{Collective decay of single-excitation eigenstates. 
    Measured waveguide decay rates \(\Gamma_{\mathrm{wg}}\) for single-excitation eigenstates of qubit arrays with size up to \(N_q=4\) under uniform engineered dissipation (\(\theta=0\), with single-qubit rate \(\gamma_{\mathrm{wg}} = 2\pi \times 15~\mathrm{kHz}\), as plotted for \(N_q=1\)). Intrinsic decay rates have been subtracted. Illustrations adjacent to each data point indicate the corresponding eigenstate wavefunction amplitudes. The highest-energy mode (\(m=1\)) exhibits superradiant decay with a rate that scales approximately linearly with \(N_q\) (dashed line).}
    \label{fig:3} 
\end{figure}

Next, we demonstrate that spectral indistinguishability of the emitted fields is required for collective decay. Keeping the qubits on resonance, we introduce a detuning \(\delta\omega_\text{mod}\) between their modulation frequencies. This detunes the emission frequencies and causes the relative phase between the two emitted fields to evolve in time, leading to alternating constructive and destructive interference. As a result, the collective decay rate becomes time dependent, oscillating between superradiant and subradiant behavior. The time-averaged decay rate therefore depends on the observation window. Here, we measure the population decay over a time window long enough for the population to fully decay, and extract an effective decay rate from a single-exponential fit. Fig.~\ref{fig:2}d shows the difference between the measured effective decay rates for initial states \(\ket{\pm}\) as a function of \(\delta\omega_\text{mod}\) and the initial modulation phase difference \(\theta\). We observe a clear suppression of the rate difference for nonzero \(\delta\omega_\text{mod}\), indicating that collective effects average out once the emissions become spectrally distinguishable. The rate difference is expected to vanish when \(\delta\omega_\text{mod}\) exceeds the relative dephasing rate of two independent qubits, \((2\gamma_\phi + \gamma_\text{wg} + \gamma_\text{int}) \approx 2\pi\times 170\,\mathrm{kHz}\), in good agreement with experiment; a detailed analysis is provided in SI Sec.\,\ref{sec:SI_extradata-sideband-detuned}.

We now extend to larger qubit arrays and study the decay dynamics of single-excitation eigenstates beyond the two-qubit case. In the presence of nearest-neighbor interactions \(J\), the single-excitation eigenstates of a degenerate \(N_q\)-site qubit lattice are standing-wave modes,
\begin{equation*}
    \ket{m} = \sum_{i=1}^{N_q} \sqrt{\frac{2}{N_q+1}} 
    \sin\!\left( \frac{\pi m i}{N_q+1} \right)\, \sigma_i^+ \ket{G},
\end{equation*}
with eigenenergies \(E_m/\hbar = 2J \cos\!\left( \frac{\pi m}{N_q+1} \right)\), where \(m=1,\dots,N_q\) and \(\ket{G}\) denotes the state with all qubits in the ground state. These modes are illustrated in Fig.\,\ref{fig:3} for \(N_q\) up to 4.
We consider the case where the engineered wavguide coupling has uniform phases on all qubits (\(\theta= \phi_{i+1}-\phi_{i}=0\), for $i=0,1,2$). The highest-energy mode \(\ket{m=1}\) has uniform phases across the lattice sites and is therefore superradiant, emitting into the waveguide with an enhanced decay rate that scales approximately linearly with the lattice size, \(\Gamma_\text{wg}^{m=1} \approx \gamma_\text{wg} N_q\). In contrast, the remaining eigenstates are subradiant with suppressed decay rates. In the ideal Dicke case, only the fully symmetric mode is bright while all orthogonal modes are perfectly dark; here, deviations from fully symmetric collective coupling lead to finite subradiant decay. We prepare these eigenstates using the same adiabatic protocol described above for the two-qubit case~\cite{Saxberg2022-tt}. The measured decay rates of the single-excitation eigenstates are shown in Fig.~\ref{fig:3}, revealing approximately linear scaling of the superradiant state with lattice size. The finite decay rates of the remaining subradiant eigenstates arise from incomplete destructive interference due to their nonuniform spatial profiles and other deviations from the ideal Dicke case discussed above, including frequency-dependent waveguide decay rates and finite spatial separation. For \(\theta=\pi\), the mode \(\ket{m=N_q}\) becomes superradiant with the same scaling, while all other states remain subradiant. The data for \(\theta=\pi\) and modeling of these deviations from the ideal Dicke model are provided in SI Sec.\,\ref{sec:SI_single_excitation_decay}.

\section{Population dynamics of superradiance}

\begin{figure*}[tbp]
    \includegraphics[width=1.9\columnwidth]{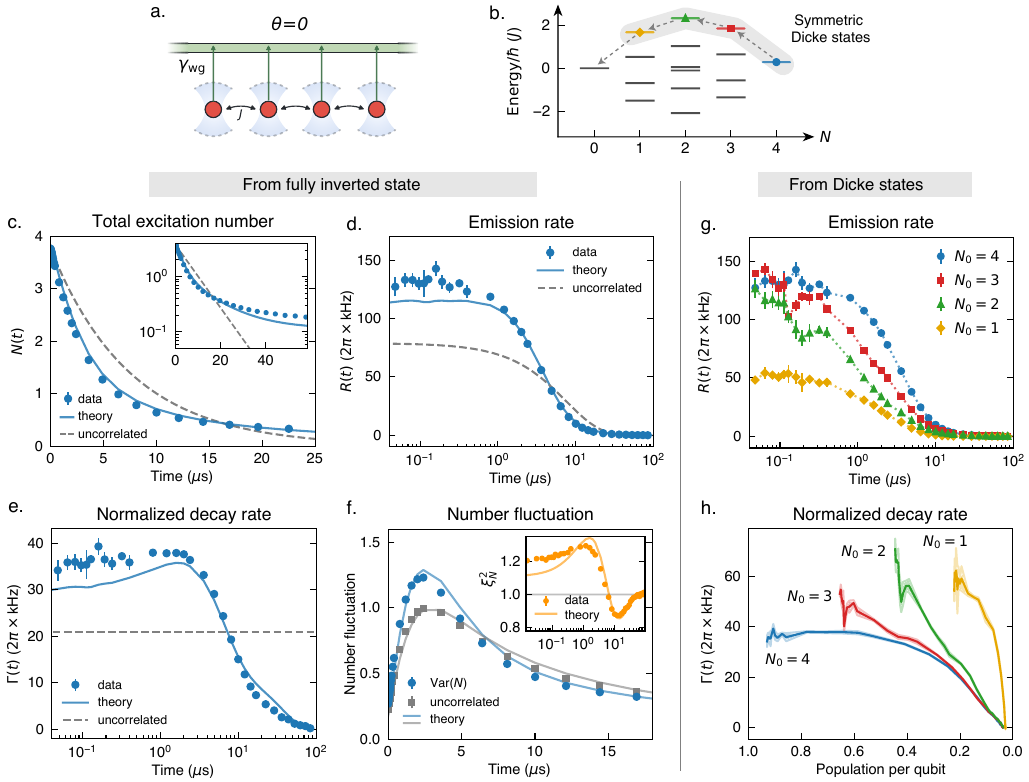}
    \caption[]{Dynamics of superradiance in a four-qubit array.
    For symmetric coupling into the waveguide (a), the superradiant manifold consists of the highest-energy eigenstates at different excitation numbers, as indicated in the many-body energy spectrum shown in the rotating frame of the qubits. These states approximate the symmetric Dicke states, while all other eigenstates are subradiant.
    For decay from the fully inverted state, we show (c) total excitation number $\langle N(t)\rangle$, (d) emission rate $R(t) = -\frac{d}{dt}\langle N(t)\rangle$, (e) instantaneous decay rate per excitation $\Gamma(t) = R(t)/\langle N(t) \rangle$, and (f) number fluctuations $\mathrm{Var}(N(t))$, and compare to expectations from uncorrelated decay (dashed lines). In (c), the inset shows the same data on a log scale. In (f), the inset shows the number-squeezing parameter $\xi_N^2$. 
    Solid lines in (c–f) are numerical simulations using measured parameters and perfect initial states.
    For decay from the Dicke-like states with initial excitation number $N_0=1,2,3,4$, we show the measured emission rate $R(t)$ in (g) and the normalized decay rate $\Gamma(t)$ versus average population per qubit in (h); the $N_0=4$ trace is the fully inverted state data from (c–f), replotted for comparison. In (g), dotted lines are guides to the eye. In (h), data and uncertainty are shown as lines and shaded regions for clarity.
    Error bars represent the standard error of the mean (s.e.m.) across 9 independently acquired datasets, each comprising $16{,}000$ single-shot measurements; uncertainties are dominated by dataset-to-dataset variations, with much smaller statistical uncertainty within each dataset. See SI Sec.\,\ref{sec:SI_rate-extraction} for details on rate extraction and error estimation. Error bars in (c,f) are smaller than the markers. 
}
    \label{fig:4} 
\end{figure*}

The decay from the fully excited state of an ensemble of emitters provides the canonical initial condition for observing superradiance. When multiple excitations are present, interference between many-body decay pathways leads to non-exponential population dynamics and, under suitable conditions, a transient enhancement of the radiated intensity at short times \cite{Masson2022-jm}.
The fully excited state of our four-qubit lattice is prepared using a time-dependent global coherent drive that sequentially adds excitations to the array via many-body Landau–Zener transitions \cite{Du2024probing}. The same protocol is used later to prepare partially excited states. For the fully inverted initial state, we achieve an average filling of $98\%$, limited by decoherence during preparation.
The engineered waveguide coupling is then turned on by ramping the flux modulation amplitude over $244\,\mathrm{ns}$ to reach a uniform decay rate $\gamma_\mathrm{wg} = 2\pi\times 15\,\mathrm{kHz}$ with uniform phases ($\theta=0$) (Fig.\,\ref{fig:4}a). After a variable evolution time $t$ under collective decay, the modulation is ramped down over another $244\,\mathrm{ns}$, the qubits are rapidly detuned from each other to suppress population transfer, and the multi-qubit state is measured.
In the absence of the ramps, abruptly turning on the qubit-resonator coupling would nonadiabatically project the initial qubit state onto a superposition of qubit-resonator dressed states, producing coherent qubit-resonator oscillations that cannot be distinguished from decay in qubit population measurements and would therefore mask the collective decay dynamics (see additional details in SI Sec.\,\ref{sec:SI_earlytime}). The ramps suppress this effect; consequently, dynamics on timescales shorter than the ramp duration are not resolved. In all measurements presented below, the evolution time $t$ excludes the ramp-on and ramp-off intervals.

The measured total excitation number $\langle N(t)\rangle$ is shown in Fig.\,\ref{fig:4}c, exhibiting a clear deviation from the exponential time dependence expected for uncorrelated, independent decay. The uncorrelated decay is characterized by a constant rate per qubit, estimated from the measured engineered and intrinsic decay rates to be $\gamma_0 = \gamma_\mathrm{wg} + \gamma_\mathrm{int} \approx 2\pi\times 21\,\mathrm{kHz}$.
In Fig.\,\ref{fig:4}d, we plot the population decay rate $R(t) = -d\langle N(t)\rangle/dt$, which closely approximates the waveguide emission rate, as waveguide decay dominates over non-radiative processes in our experiments. Starting from the fully inverted state, the emission is expected to be initially uncorrelated, corresponding to a rate $\gamma_0 \langle N(t)\rangle$ at very short times, although this regime is not resolved experimentally because of the finite ramp duration. As emission proceeds under symmetric coupling to the waveguide ($\theta=0$), spontaneous emission seeds the buildup of correlations, leading to enhanced collective decay through superradiant many-body eigenstates. 
In the measured dynamics, the extracted emission rate remains approximately constant at \(R(t)\approx 2\pi\times130\,\mathrm{kHz}\) during the initial stage of the decay (\(t\lesssim200\,\mathrm{ns}\)). It then gradually decreases, while remaining above the uncorrelated expectation \(\gamma_0\langle N(t)\rangle\) until \(t\approx5\,\mu\mathrm{s}\).
We also define the instantaneous decay rate per excitation $\Gamma(t) = R(t)/\langle N(t) \rangle$, shown in Fig.\,\ref{fig:4}e. In the remainder of the paper, we identify superradiance (subradiance) in the many-body setting with the condition $\Gamma(t) > \gamma_0$ ($\Gamma(t) < \gamma_0$). 
The initial rapid population depletion during superradiance is followed by a long-lived tail, in which population accumulates in slow-decaying subradiant states, persisting over $t\sim10-100\,\mu\mathrm{s}$. The crossover $\Gamma(t)=\gamma_0$ marks the transition from superradiant to subradiant dynamics and occurs at $t\approx 7.5\,\mu\mathrm{s}$ ($\gamma_0 t \approx 1.0$).
The measurements are in good agreement with numerical simulations using independently calibrated parameters and assuming a perfectly inverted initial state. 
The slightly enhanced measured emission and decay rates at early times likely arise from imperfections in the initial-state preparation that seed the collective decay dynamics, as well as residual flux-control crosstalk in the device that distorts the modulation amplitudes and induces weak time dependence in the engineered decay rates. Details of the numerical model are provided in SI Sec.\,\ref{sec:SI_full-model}.

In the absence of coherent interactions between identical emitters, collective decay is fully determined by the spatial structure of the coupling to the dissipative environment, which directly defines the superradiant and subradiant modes. In our strongly interacting qubit array, however, the coherent Hamiltonian determines the many-body eigenstates. In the regime \(J \gg \gamma_\mathrm{wg}\), transitions within the many-body spectrum become spectrally resolved, and collective decay proceeds through distinct decay channels rather than through a single collective mode. The resulting dynamics therefore arise from a cascade through the interacting many-body spectrum, with transition rates set jointly by the frequency-dependent waveguide coupling and the transition matrix elements between eigenstates. For example, decay from the fully excited $N=4$ state proceeds through four spectrally distinguishable transitions into the $N=3$ manifold. The highest-energy $N=3$ eigenstate has uniform phases and exhibits the strongest transition rate for $\theta=0$. The remaining three states are subradiant with finite transition rates because of finite overlap with the symmetric decay channel and the energy dependence of $\gamma_\mathrm{wg}$. The subsequent dynamics continue through a cascade down the many-body spectrum. An effective model for calculating the many-body decay rates is described in SI Sec.\,\ref{sec:SI_manybody-rates}.

In Fig.\,\ref{fig:4}f, we extract fluctuations in $N$ from the measured qubit-resolved populations. The population variance is
$\mathrm{Var}(N) = \langle (\Delta N)^2 \rangle = \sum_i p_i(1-p_i) + 2\sum_{i<j} C_{ij}$,
where $C_{ij}=\langle n_i n_j\rangle-\langle n_i\rangle\langle n_j\rangle$ is the connected two-point density correlation between qubit $i$ and $j$. The term $\sum_i p_i(1-p_i)$ sets the shot-noise (binomial) baseline for independent decay, while $C_{ij}$ captures correlations arising from collective dynamics. We quantify departures from the shot-noise baseline using the number-squeezing parameter $\xi_N^2=\mathrm{Var}(N)/\sum_i p_i(1-p_i)$.
For decay from the fully inverted state, $\mathrm{Var}(N)$ and $\xi_N^2$ are shown in Fig.\,\ref{fig:4}f. During the superradiant regime, $\xi_N^2>1$, indicating enhanced number fluctuations relative to the uncorrelated limit. At later times, $\xi_N^2<1$ as population accumulates in subradiant states with low excitation number. The transition at $t\approx 7.5\,\mu\mathrm{s}$ coincides with the onset of subradiant dynamics identified from $\Gamma(t)$, reflecting the redistribution of population from bright collective states to subradiant states.
Compared to numerical simulations, imperfections in the initial state slightly enhance the initial $\xi_N^2$ and modify its subsequent dynamics during superradiance.

\begin{figure*}[!tbp]
    \includegraphics[width=1.5\columnwidth]{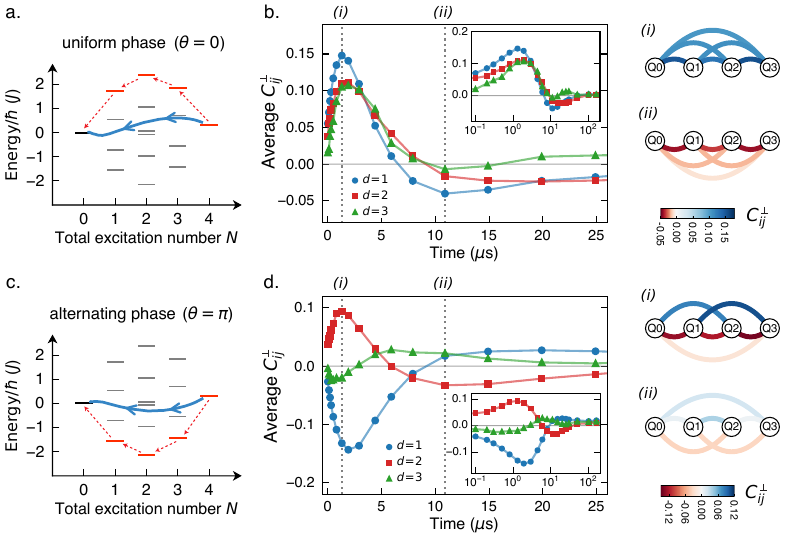}
    \caption{
    Programmable correlations from collective decay. 
    (a,\,c) Energy spectra showing the dominant decay pathways under engineered decay with uniform and alternating phases. Superradiant (subradiant) states are shown in red (grey).
    Solid blue curves show the many-body energy $E(t)$ versus $N(t)$ during decay, extracted from correlation measurements; the arrows indicate the time direction.
    (b) Connected symmetric transverse correlations $C_{ij}^{\perp}=(C_{ij}^{xx}+C_{ij}^{yy})/2$ measured during decay from the fully inverted state for uniform collective decay ($\theta=0$), averaged over qubit pairs with the same separation $d$. Positive correlations develop during the superradiant dynamics, followed by sign reversal during the subradiant dynamics. The inset shows the same data on a log time scale. Pair-resolved correlation matrices at representative times during the superradiant (i) and subradiant (ii) stages of the dynamics. 
    (d) Correlations for decay with alternating phases ($\theta=\pi$), exhibiting a spatially alternating pattern with negative nearest-neighbor and positive next-nearest-neighbor correlations, consistent with the engineered phase structure of the collective decay. Solid lines in (b,d) are guides to the eye. Uncertainties in $C_{ij}^{\perp}$ are dominated by systematic errors estimated to be $\approx 0.005$, smaller than the data markers.
}
    \label{fig:5} 
\end{figure*}

Furthermore, we investigate the dynamics of partially excited states, where collective decay is influenced by pre-existing correlations between the qubits. In the ideal Dicke case, symmetric Dicke states constitute the superradiant manifold and possess the symmetric correlations required for collective emission, leading to an enhanced decay rate at $t=0$. This contrasts with the fully inverted state, which, although part of the symmetric manifold, exhibits an uncorrelated decay rate at $t=0$. For the ideal Dicke model, the total emission rate of a Dicke state with $N_0$ excitations is $\gamma_0N_0(N_q-N_0+1)$; the corresponding normalized decay rate therefore scales as $\gamma_0(N_q-N_0+1)$, increasing linearly with decreasing $N_0$.
Using the same coherent drive protocol, we prepare multi-qubit states with a fixed number of excitations $N_0=1,2,3,4$ by populating the highest-energy eigenstates in each excitation-number sector of the coupled qubit lattice, as illustrated in the many-body spectrum in Fig.\,\ref{fig:4}b. For $\theta=0$, these states form the symmetric superradiant manifold that dominates the early-time decay. In a finite lattice, they closely resemble the symmetric Dicke states, with overlap $\geq93\%$ for $N_q=4$ (see SI Sec.\,\ref{sec:SI_state-prep}). In Fig.\,\ref{fig:4}g, we show the measured emission rate $R(t)$ for these Dicke-like states. In Fig.\,\ref{fig:4}h, we show the measured normalized decay rate as a function of the average population per qubit. Although residual short-time oscillations and the finite ramp complicate a quantitative comparison, the overall behavior of $\Gamma(t)$ remains consistent with the Dicke-model prediction: the average initial decay rate increases with decreasing $N_0$.

\section{Programmable quantum correlations}

Capitalizing on the site-resolved measurement capabilities of superconducting circuits, we next probe qubit--qubit correlations underlying collective radiation and demonstrate programmable control over their spatial structure.
The number fluctuations measured in Fig.\,\ref{fig:4}f are directly related to pair correlations of excitation number $\langle n_i n_j\rangle$, or equivalently to longitudinal spin correlations $\langle \sigma_i^z\sigma_j^z\rangle$ through $n_i=(1+\sigma_i^z)/2$. 
In contrast, the transverse correlations $\langle \sigma_i^x \sigma_j^x\rangle$ and $\langle \sigma_i^y \sigma_j^y\rangle$ probe coherent phase correlations between qubits. Their buildup reflects the formation of collective dipole correlations during superradiant decay and the emergence of spatially structured subradiant states. 
Here, we focus on decay from the fully inverted initial state and measure the connected symmetric transverse correlation
$C_{ij}^{\perp} = (C_{ij}^{xx}+C_{ij}^{yy})/2$,
where
$C_{ij}^{xx} =
\langle \sigma_i^x \sigma_j^x\rangle -
\langle \sigma_i^x\rangle
\langle \sigma_j^x\rangle$ and similarly for $C_{ij}^{yy}$.
For the fully inverted initial state, the dynamics are rotationally symmetric in the $xy$-plane. We therefore expect $\langle \sigma_i^x\sigma_j^x\rangle=\langle \sigma_i^y\sigma_j^y\rangle$, and $\langle \sigma_i^x\rangle=\langle \sigma_i^y\rangle=0$ throughout the observed dynamics.

We begin by studying collective decay with uniform phases, $\theta=0$. We measure all pair correlations by performing suitable single-qubit Pauli rotations prior to projective measurement in the number basis.
Fig.\,\ref{fig:5}b shows the measured $C_{ij}^{\perp}(t)$, averaged over qubit pairs with the same separation $d=|i-j|$, for $d=1,2,3$. 
During the superradiant dynamics, positive transverse correlations develop across all distances, with the nearest-neighbor correlation reaching a maximum value of $C_{ij}^{\perp}\approx 0.15$ at $t=1.3\,\mu$s. The positive sign is consistent with the buildup of an in-phase collective dipole for $\theta=0$, where superradiant decay occurs through symmetric eigenstates. This is followed by longer-lived subradiant dynamics in which the correlations reverse sign for both nearest-neighbor ($d=1$) and next-nearest-neighbor ($d=2$) pairs, reflecting the spatial phase structure of the subradiant states. The correlations at $d=3$ show reduced contrast and are more sensitive to finite-size effects and residual disorder in the low-population subradiant dynamics. At late times, all correlations decay to zero as the system approaches the ground state. Pair-resolved correlations at representative times during the superradiant ($t=1.3\,\mu$s) and subradiant ($t=11\,\mu$s) stages of the dynamics are shown in Fig.\,\ref{fig:5}b. The full time-dependent correlation data are presented in SI Sec.\,\ref{sec:SI_extradata-correlation}.

The measured pair correlations also enable reconstruction of the instantaneous energy of the many-body state. Using the measured transverse correlations, we evaluate the expectation value of the qubit-lattice Hamiltonian including nearest-neighbor and next-nearest-neighbor interactions, while neglecting the much smaller distance-three interaction. We additionally include a small density-dependent energy correction arising from finite transmon anharmonicity \(U\), which is also incorporated into the eigenenergies shown in the spectra (see SI Sec.\ref{sec:SI_extradata-correlation}). We show the resulting energy \(E(t)\) versus excitation number \(N(t)\) in Fig.\,\ref{fig:5}a, revealing a trajectory that clearly deviates from the highest-energy superradiant states and demonstrating the dynamical evolution into subradiant manifolds during the decay process.

When the engineered decay has alternating phases ($\theta=\pi$), neighboring qubits emit out of phase into the waveguide, which qualitatively alters the correlation pattern. The lowest-energy eigenstates at each excitation number now form the superradiant manifold, while the remaining eigenstates become subradiant. This difference is illustrated in the energy spectra in Fig.\,\ref{fig:5}c. Due to the approximate inversion symmetry of the spectra in our system, the resulting population dynamics remain similar to those for $\theta=0$; see additional data in SI Sec.\,\ref{sec:SI_extradata-out-of-phase}. 

As shown in Fig.\,\ref{fig:5}d, the engineered decay with $\theta=\pi$ produces a spatially alternating correlation pattern: during superradiance, nearest-neighbor transverse correlations ($d=1$) become negative with a peak magnitude similar to the $\theta=0$ case, while next-nearest-neighbor correlations ($d=2$) remain positive. This follows directly from the imposed phase structure: nearest neighbors acquire a relative phase of $\pi$, while next-nearest neighbors acquire a relative phase of $2\pi$. Similarly, correlations at $d=3$ correspond to a relative phase of $3\pi$ and therefore reverse sign again. As in the uniform case, the $d=3$ correlations are more strongly affected by finite-size effects. We again observe a reversal of the correlations as the system evolves from superradiant to subradiant dynamics. Representative pair-resolved correlations during the superradiant ($t=1.3\,\mu$s) and subradiant ($t=11\,\mu$s) stages of the dynamics are shown in Fig.\,\ref{fig:5}d. We similarly reconstruct the many-body energy from the measured pair correlations and plot the resulting average trajectory in Fig.\,\ref{fig:5}c.

Overall, these measurements reveal how engineered collective dissipation dynamically generates and reshapes many-body correlations in interacting qubit arrays. These results establish programmable dissipation as a powerful resource for controlling correlated quantum dynamics in many-body quantum systems.

\section{Outlook}

Our experiments provide microscopic access to the populations and correlations of individual emitters throughout the collective decay process. Extending these measurements to the emitted microwave fields and their coherent properties \cite{Bozyigit2011-qn} would offer further insight into the role of entanglement in collective emission from interacting many-body systems. Joint measurements of the qubits and emitted photons could further elucidate the buildup of emitter-photon correlations and the emergence of collective phases, including recently explored superradiant phase transitions \cite{Ferioli2023-nc, Song2025-lf}. The high tunability of superconducting circuits also enables systematic studies of collective emission in the presence of controlled disorder, tunable interactions, and engineered decoherence.

Beyond simple waveguide environments, superconducting circuits support a wide range of engineered photonic reservoirs, including photonic bandgap metamaterials \cite{Liu2017-pb} and topological photonic lattices \cite{Kim2021-fi}. Such environments modify both the photonic density of states and the spatial structure of emitter-environment coupling, allowing exploration of non-Markovian collective dynamics and long-range photon-mediated interactions. Engineered non-reciprocal couplings \cite{Joshi2023-kc} and artificial giant atoms \cite{Kannan2020-tw} further enable directional emission that can give rise to exotic multi-photon bound states \cite{Mahmoodian2020-cz}.

Programmable collective dissipation also creates new opportunities for dissipative quantum state engineering. Subradiant states in emitter arrays could serve as long-lived quantum memories \cite{Asenjo-Garcia2017-tb}, while the interplay of coherent driving, incoherent pumping, and collective dissipation could be utilized to stabilize correlated many-body states \cite{Scarlatella2025-ka, Mann2025-fo}. Dynamically tunable waveguide emission further enables controlled generation of propagating photonic states, including entangled photonic states for quantum communication and networking \cite{PhysRevLett.115.163603, Rubies-Bigorda2025-ap}. More broadly, superconducting qubit lattices with programmable collective coupling provide a versatile platform for analog quantum simulation of open many-body quantum systems \cite{Zhang2023-fg, Fayard2021-mv}, and could enable applications in quantum-enhanced sensing \cite{Paulisch2019-ar}.

\section*{Acknowledgments}

The authors thank Francis Robicheaux and Jonathan Simon for helpful feedback on the manuscript.

This work was supported by grants from the Air Force Office of Scientific Research (award number FA9550-23-1-0491) and the National Science Foundation (award number DMR-2145323). Part of this material is based upon work supported by the U.S. Department of Energy, Office of Science, National Quantum Information Science Research Centers, Quantum Science Center.



\input{main.bbl}

\makeatletter
\newcounter{mainbibcount}
\setcounter{mainbibcount}{\value{NAT@ctr}}
\makeatother

\clearpage

\let\addcontentsline\oldaddcontentsline

\input{supplementary.tex}

\end{document}

%% file: main.bbl
%

%% file: supplementary.tex
\clearpage
\setcounter{page}{1}
\renewcommand{\thepage}{S\arabic{page}}

\renewcommand{\thetable}{S\arabic{table}}
\renewcommand{\theequation}{S\arabic{equation}}
\renewcommand{\thefigure}{S\arabic{figure}}
\setcounter{equation}{0}
\setcounter{figure}{0}
\setcounter{secnumdepth}{3}

\newcommand{\mi}{\mathrm{i}}
\newcommand{\dif}{\mathrm{d}}

\renewcommand{\appendix}{\par
  \setcounter{section}{0}
  \setcounter{subsection}{0}
  \setcounter{subsubsection}{0}
  \gdef\thesection{\Alph{section}}
  \gdef\thesubsection{\Alph{section}.\arabic{subsection}}
  \gdef\thesubsubsection{\Alph{section}.\arabic{subsection}.\roman{subsubsection}}
}

\onecolumngrid
\appendix

\begin{center} 
    \uppercase{\textbf {Supplementary Information}}\\
    \vspace{0.2in}
    {\large \textbf \MyTitle} \\[0.5em]
\end{center}





\makeatletter
\renewcommand\p@subsection{}
\makeatother

\begingroup
\makeatletter

\let\oldnumberline\numberline
\renewcommand*\numberline[1]{\oldnumberline{#1.}}

\renewcommand*\l@section[2]{%
  \addvspace{0.4\baselineskip}
  \@dottedtocline{1}{0em}{1.6em}{#1}{#2}%
}
\renewcommand*\l@subsection{\@dottedtocline{2}{1.3em}{2.4em}}
\renewcommand*\l@subsubsection{\@dottedtocline{3}{2.8em}{3.2em}}

\makeatother
\tableofcontents
\endgroup

\section{Device parameters} 
\label{sec:SI_device-param}

The superconducting qubit device used in this work is identical to that in previous studies \cite{Du2024probing, Du2025-iw}. Detailed information on device fabrication, cryogenic wiring, room-temperature measurement setup, and procedures for device characterization and tuning can be found in the Supplemental Material of Ref.~\cite{Du2024probing}.
Table \ref{tab:device_parameters} lists the relevant experimental parameters used in this work, which are also used in the numerical simulations presented in this supplementary material.

\begin{table}[ht]
\centering
\begin{tabular}{lcccc}
\hline\hline
\textbf{Qubits} & Q0 & Q1 & Q2 & Q3 \\
\hline
Frequency $\omega_\text{q}/2\pi$ (MHz)            & 4600 & 4600 & 4600 & 4600 \\
Anharmonicity $U_2/2\pi$(MHz)              & $-245.3$ & $-245.6$ & $-245.5$ & $-245.7$ \\
$T_1$ ($\mu$s)                   & 33.1 & 30.6 & 31.0 & 43.1 \\
$T_2^*$ ($\mu$s)                 & 1.9  & 1.8  & 1.1  & 1.8 \\
Thermal population   & 0.05 & 0.05 & 0.05  & 0.05 \\
\hline
Nearest-neighbor $J_{i,i+1}/2\pi$ (MHz)  & 5.86 & 5.89 & 5.86  & \\
$J_{i,i+2}/2\pi$ (MHz)   & 0.6 & 0.6 &   &  \\
\hline
\noalign{\vskip 1.0ex}
\textbf{Resonators} & R0 & R1 & R2 & R3 \\
\hline
Frequency $\omega_\text{r}/2\pi$ (MHz)        & 6323.2 & 6277.8 & 6231.2 & 6180.2 \\
Qubit-res. coupling $g/2\pi$ (MHz)               & 68.5   & 65.7   & 63.7   & 69.2 \\
Linewidth $\kappa_\text{r}/2\pi$ (MHz)         & 1.54   & 1.23   & 1.61   & 1.40 \\
Dispersive shift $2\chi/2\pi$ (MHz)   & 0.678  & 0.656  & 0.651  & 0.820 \\
Thermal population   & 0.025 & 0.025 & 0.025 & 0.025 \\
\hline\hline
\end{tabular}
\caption{Qubit and resonator parameters.}
\label{tab:device_parameters}
\end{table}

The nearest-neighbor qubit--qubit coupling $J_{i,i+1}$ is measured by observing the exchange dynamics of a single excitation initialized in one of two resonantly coupled qubits. The next-nearest-neighbor tunneling $J_{i,i+2}$ is estimated from finite-element microwave simulations of the device and is consistent with the measured spectra of the coupled qubit array. While all experiments in this work use only the lowest two levels of the transmon qubits, the unoccupied higher levels still contribute to frequency shifts of the coupled lattice modes, especially for states with higher total excitation number. This is visible qualitatively in the qubit spectra in Figs.\,\ref{fig:4} and \ref{fig:5}, and is included in the numerical simulations.

Individual resonator parameters are extracted from ``Chi-Kappa-Power'' measurements \cite{Sank2025-qf}.
The finite separation between the resonator couplers along the waveguide can be estimated by driving the qubits through the transmission line and extracting the relative phase between the drive fields seen by the qubits. This was previously measured at 4.5\,GHz to correspond to a relative phase of $7^\circ$ (see \cite{Du2024probing} for details). At the emission frequency of approximately 6.25\,GHz used in this work, this corresponds to a relative phase of $10^\circ$, or equivalently $d \approx 0.03 \lambda$.

\subsection{Intrinsic qubit relaxation}
\label{sec:SI_intrinsic_decay}
We consider three contributions to qubit relaxation: intrinsic non-radiative relaxation $\gamma_\text{nr}$, intrinsic Purcell decay into the waveguide $\gamma_\text{purc}$ (through the resonators or other potential package modes), and the engineered Purcell decay through the resonator into the waveguide $\gamma_\text{wg}$. Both the intrinsic and engineered Purcell decay can result in interference in the waveguide and, therefore, collectively enhanced or suppressed effective rates. However, these two decay channels are independent from each other since they correspond to different emission energies ($\omega_\text{q}$ for the intrinsic Purcell decay and $\omega_\text{q}' = \omega_\text{q}+\omega_\text{mod}$ for the engineered waveguide decay).
The listed $T_1$ in Table \ref{tab:device_parameters} is the measured relaxation rate of individual qubits at the idle lattice frequency (4.6 GHz) while the other qubits are far detuned. Hence, it includes contributions from both non-radiative relaxation $\gamma_\text{nr}$ and the intrinsic Purcell decay $\gamma_\text{purc}$. From the measured individual qubit $T_1 \approx (30-40)\,\mu$s, we get $\gamma_\text{nr} + \gamma_\text{purc} \approx 1/T_1 \approx 2\pi\times (4-5)$\,kHz.
The intrinsic Purcell decay can be calculated, assuming dominant contribution from the fundamental mode of the resonator, to be $\gamma_\text{purc} = \kappa g^2/\Delta^2 \approx 2\pi\times(2-2.5)$\,kHz, which in turn provides an estimate for $\gamma_\text{nr} \approx 2\pi\times(2-2.5)$\,kHz. 

\begin{figure}[!tbp]
    \centering
    \includegraphics[width=0.8\columnwidth]{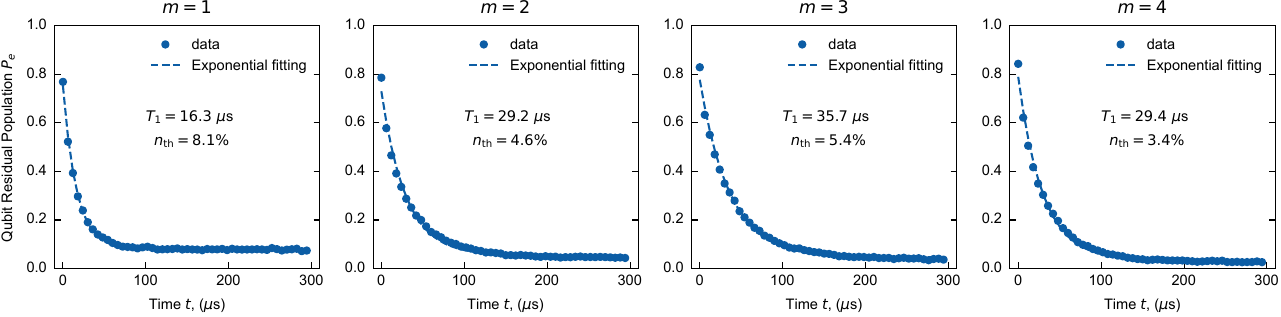}
    \caption[]{Intrinsic Purcell decay of single-excitation eigenstate in the 4 qubit array. See also Fig.~\ref{fig:phase_vector_exp}(f).}
    \label{fig:int_col_decay} 
\end{figure}

In Fig.\,\ref{fig:int_col_decay}, we show the intrinsic decay rate of the single excitation eigenstates with 4 qubits. This follows the same experimental sequence as the ones presented in Fig.\,\ref{fig:3}, but in the absence of the engineered waveguide coupling (no flux modulation).
The observed decay rates are due to both non-radiative decay and the Purcell-induced collective decay. We measure the decay rates of eigenstates $m$ to be $2\pi\times\{9.8, 5.4, 4.5, 5.4\}\,$kHz, where $m$ indexes the eigenstates in descending order of eigenenergy. The first entry corresponds to the highest-energy, uniform-phase eigenstate, which exhibits an enhanced collective decay rate arising from intrinsic Purcell decay. We expect the uniform phase state to decay at $\approx 4\gamma_\text{purc} + \gamma_\text{nr}$, which is close to the measured value. The other three subradiant eigenstates should decay at roughly $\gamma_\text{nr}$. The observed linewidths are a few kHz larger than this simple expectation, likely due to inhomogeneity in the intrinsic Purcell decay rates and other imperfections that break ideal symmetric collective coupling, including finite $J$ and finite $d$. 

\subsection{Dephasing}
\label{sec:SI_dephasing}
We measure single-qubit dephasing $\Gamma_\phi = 1/T_2^* \approx 2\pi \times 75\,\text{kHz}$ from Ramsey experiments. This dephasing is limited by flux noise from the room-temperature electronics used to bias the qubit frequencies, with a typical $1/f$-type noise spectrum. This dephasing rate exceeds the engineered collective decay rate $\gamma_\text{wg} = 2\pi\times 15\,\text{kHz}$.

In our experiments, the strong coherent interaction $J$ between the qubits suppresses the effect of local dephasing. Consider the example of two resonantly coupled qubits. In the presence of a small detuning $\delta$, for example due to local dephasing, the energy eigenvalues are $\pm\sqrt{J^{2}+(\delta/2)^{2}}\approx\pm\bigl[J+\delta^{2}/(8J)\bigr]$. Therefore, the effective linewidth of the single-particle eigenstates due to dephasing is suppressed to $\Gamma_\phi^\text{eff} \approx \Gamma_\phi^{2}/(4J) \approx 2\pi\times 0.2\,\text{kHz}$. This effect has previously been used in dual-rail qubits implemented with two coupled transmons \cite{Levine2024-yd_truncate}. Hence, the engineered collective decay in our experiment dominates the effective dephasing and intrinsic relaxation, leading to the observed super- and subradiant emission. Below, in Sec.\,\ref{sec:SI_extradata-qubits-detuned}, we measure collective decay from two detuned qubits, where the effect of local dephasing is restored.

\subsection{Thermal population}
\label{sec:SI_nthermal}

Qubit thermal populations are measured by idling all qubits in the degenerate coupled lattice until equilibrium is reached, then rapidly detuning each qubit and measuring its excited-state population using an $ef$-Rabi experiment. We use a measurement-based reset protocol to suppress thermal population in the initial state (see Sec.\,\ref{sec:SI_state-prep}).

To measure the resonator thermal population, we apply a resonant red-sideband flux modulation for a controlled duration to swap the resonator thermal population into the qubit, and then measure the qubit state. The sideband dynamics, including resonator decay, are numerically modeled to extract the resonator thermal population ($n^{\text{R}}_{\mathrm{th}}\approx 2.5~\%$). This provides an estimate of the effective waveguide temperature near the resonator frequencies.

The effective temperature of the intrinsic collective decay channels can be extracted directly from the decay dynamics. From the long-time qubit populations in Fig.\,\ref{fig:int_col_decay}, we extract an effective thermal population of approximately $8.1\%$ for the environment at around $4.6$\,GHz that couples to the symmetric mode via the intrinsic Purcell decay and $3.4$--$5.4\%$ for the other modes.

The effects of these thermal populations on the collective decay dynamics and correlations are modeled and discussed in Sec.\,\ref{sec:SI_modeling}.

\section{Experimental sequence}
\label{sec:SI_expt-seq}

\subsection{State preparation}
\label{sec:SI_state-prep}

\subsubsection{Multi-qubit reset via heralding measurement}

To mitigate the impact of qubit thermal population on state preparation and readout calibration, we employ a heralding measurement to prepare the true ground state. We perform frequency-multiplexed readout at the lattice location, where the qubits are resonantly coupled, using a typical readout duration of $1.5\,\mu\text{s}$. Although resolving individual Fock states in this configuration is challenging, the ground state can be clearly distinguished from all other states in the multi-tone quadrature signal. Using the control hardware's real-time conditional logic (Quantum Machines OPX+), we determine whether the qubits are in the ground state based on calibrated thresholds. If the heralding condition is not satisfied, indicating a high probability that at least one qubit is excited, we wait $300\,\mu\text{s}$ for the system to re-equilibrate and repeat the heralding sequence until successful. After successful heralding, we wait an additional $1.6\,\mu\text{s}$ to deplete photons in the readout resonators before applying the main experimental sequence. To minimize residual thermal population, we use aggressive state-discrimination thresholds, resulting in a heralding success probability of approximately $8\%$. Using $ef$-Rabi experiments, we verify that the residual thermal population in the heralded state is below $1\%$ per qubit. In future experiments, the heralding efficiency and repetition rate could be significantly improved using active reset and real-time feedback.

\subsubsection{Single excitation eigenstates}

Experiments in Fig.\,\ref{fig:3} use initial states corresponding to single-excitation eigenstates of the coupled qubit array. These are prepared using a disorder-assisted adiabatic scheme \cite{Saxberg2022-tt}. In the 4-qubit case, starting from the heralded ground state, the qubits are rapidly detuned within $2\,\text{ns}$ to $\{233, 121, 174, 64.5\} \times 2\pi\,\text{MHz}$, and the qubit with the $m$th highest frequency is excited with a resonant $\pi$ pulse applied through the readout transmission line. The detunings are then adiabatically ramped to zero over $420\,\text{ns}$ to prepare the $m$th quasi-momentum eigenstate. The ramp is chosen to be sufficiently slow to suppress nonadiabatic transitions, such that the remaining infidelity is primarily incoherent. Due to decoherence during the ramp, the prepared single-excitation eigenstates have a typical fidelity of approximately $90\%$.

\subsubsection{Dicke-like states at different lattice filling}

The experiments in Fig.\,\ref{fig:4}(g-h) start from the highest-energy eigenstate in each excitation-number manifold. These states have fixed excitation number and uniform wavefunction phases across the qubits. We denote the highest-energy eigenstate of a lattice with $N_q$ qubits and excitation number $N$ by $\ket{\psi_{N_q}^{(N)}}$.

We prepare these states using a time-dependent global drive that sequentially adds excitations through adiabatic many-body Landau-Zener transitions \cite{Du2024probing, Du2025-iw}. Starting from the ground state of the degenerate qubit array, a global drive is applied through the readout transmission line and ramped up over $300\,\text{ns}$ ($100\,\text{ns}$ for $N = 4$) to Rabi rates of $\{2.1, 3.1, 2.1, 4.5\} \times 2\pi\,\text{MHz}$ for filling numbers $N = 1$ through $4$, respectively, with an initial detuning of $+2\pi \times 30\,\text{MHz}$. The drive frequency is then swept at rates of $\{-74.6, -102.7, -142.0, -416.7\} \times 2\pi\,\text{MHz}$ to final detunings of $\{7.6, 0.8, -5.8, -30\} \times 2\pi\,\text{MHz}$, after which the drive amplitude is ramped down over $300\,\text{ns}$ ($100\,\text{ns}$ for $N = 4$). The sweep parameters are optimized experimentally for each filling.
Numerical simulations estimate target-state populations between $85\%$ and $92\%$ for $N = 1$ through $4$. This is consistent with the measured fidelity of $92\%$ for the fully inverted state ($N = 4$), corresponding to an average filling of $98\%$. Simulations indicate that decoherence contributes approximately $4.0\%$ to the total infidelity, with the remainder arising from diabatic excitations during preparation.

\begin{figure}[!htbp]
    \centering
    \includegraphics[width=0.4\columnwidth]{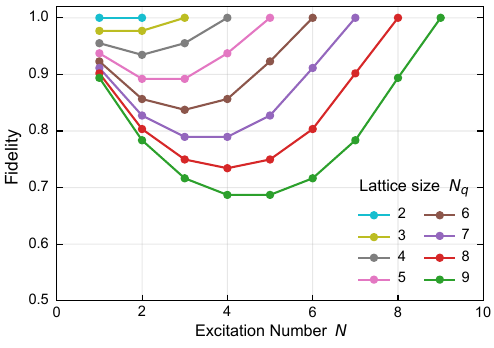}
    \caption[]{Fidelity between the highest-energy lattice eigenstates and the symmetric Dicke states.}
    \label{fig:overlap_to_Dicke_states} 
\end{figure}

In finite lattices, these states closely resemble the symmetric Dicke states
\[
\ket{D_{N_q}^{(N)}} = \binom{N_q}{N}^{-1/2}
\sum_{\substack{\mathbf{n}\in\{0,1\}^{N_q}\\ \sum_i n_i = N}}
\ket{n_1 n_2 \cdots n_{N_q}}.
\]
Figure~\ref{fig:overlap_to_Dicke_states} shows the fidelity, defined here as the population overlap between the highest energy lattice eigenstates and the corresponding symmetric Dicke states 
$F^{(N)}_{N_q} =
\left|\braket{D_{N_q}^{(N)}}{\psi_{N_q}^{(N)}}\right|^2$,
for different lattice sizes. For our experimental case of $N_q = 4$, the overlap exceeds $93\%$, ignoring preparation errors.

\subsection{Evolution under engineered collective decay}
\label{sec:SI_eng-decay}

\subsubsection{Calibration of waveguide coupling rate and relative phase}

\begin{figure}[!htbp]
    \centering
    \includegraphics[width=1\columnwidth]{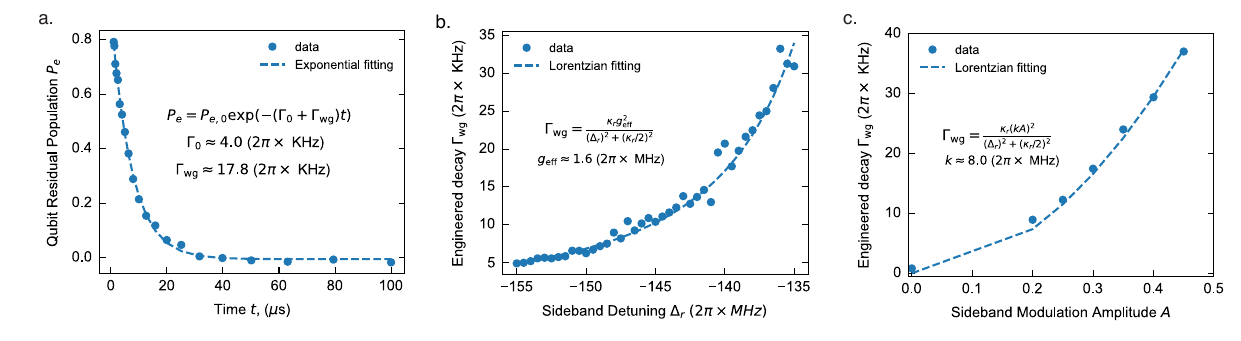}
    \caption{
    Calibration of qubit-waveguide coupling rates. (a) Engineered single-qubit decay. We couple a single qubit to the waveguide via flux modulation and extract the engineered decay rate $\Gamma_{\mathrm{wg}}$ from $T_1$ decay measurements. (b) Dependence of $\Gamma_{\mathrm{wg}}$ on the sideband modulation detuning $\Delta_r$, fitted to a Lorentzian profile. (c) Dependence of $\Gamma_{\mathrm{wg}}$ on the sideband modulation amplitude $A$, also showing a Lorentzian fit. For the main experiments, we select a sideband modulation amplitude that yields $\Gamma_{\mathrm{wg}} \approx 15 \times 2\pi$ kHz.
    }
    \label{fig:calibration_sb_combined} 
\end{figure}

The flux modulation setup, parameters, and achievable ranges for the sideband coupling $g_{\mathrm{eff}}$ are detailed in Ref.~\cite{Du2025-iw}. We induce engineered collective decay using a flux-modulation frequency of roughly $2\pi\times 1.657$ GHz, placing the modulated qubit sideband near the average resonator frequency. We choose modulation powers to yield a uniform engineered decay rate of $\Gamma_{\mathrm{wg}} \approx 2\pi\times 15$ kHz.
We calibrate the sideband amplitudes by parking each qubit at the lattice frequency while detuning the remaining qubits, and measure the red-sideband-induced relaxation time of the target qubit as a function of the flux-modulation amplitude. Figure~\ref{fig:calibration_sb_combined} illustrates the typical calibration process for the qubit-waveguide coupling rates.

\textit{Calibration of relative phase:}
The sideband modulation pulses applied to different qubits acquire different phases due to variations in the lengths of the microwave control and on-chip flux lines. For each neighboring qubit pair, we prepare symmetric and antisymmetric single-excitation states, apply engineered decay with matched amplitudes, and sweep the relative phase between the two modulation tones. The phase that maximizes $(\Gamma_+ - \Gamma_-)$ is defined as the uniform-phase condition ($\theta = 0$).

\subsubsection{Smooth ramping of flux modulation}

In the experiments, the flux modulation is ramped smoothly to the target amplitude in $244$\,ns. This soft turn-on is to minimize coherent oscillations in the qubit population at short times: A sudden quench from the bare qubit states into the dressed states of the qubit-resonator system induced coherent oscillations. In addition, imperfect compensation of the modulation-induced qubit frequency shift can result in quench dynamics leading to population oscillation between qubits, which can also manifest in the total excitation number when there is a slight inhomogeneity in the waveguide decay rates for different qubits.
In our experiments, the first effect is dominant. In the numerical modeling presented in Sec.~\ref{sec:SI_earlytime}, we compare the dynamics of the qubit excitation number with the total excitation across both the qubits and resonators to demonstrate that the observed oscillations in the measured decay rates at early times are dominated by coherent exchange between the qubits and the resonators.

The flux modulation also induces an effective qubit frequency shift, typically $\lesssim 2$ MHz, due to the non-linear frequency response of the qubit to flux and the off-resonance charge drive. We characterize these shifts using Ramsey experiments while the modulation is active. We then compensate for them by applying DC offsets to the flux pulses, ensuring the qubits remain on resonance during the collective decay.

\subsection{Readout and state characterization}
\label{sec:SI_readout}

\subsubsection{Multiplexed readout}

Multi-qubit Fock states (the computational basis of the qubits) are used to calibrate the readout signal. Starting from the heralded ground state, we rapidly detune all nearest-neighbor qubits by more than 400 MHz within $2$ ns. We wait for $80$~ns for the flux pulses to settle before sequentially applying the resonant $\pi$ pulses, implemented as a DRAG-Gaussian pulse with a length of $80$~ns.

A frequency-multiplexed readout pulse is then applied for a typical duration of 1.5 $\mu$s. The obtained multi-qubit I/Q signals are used to train a support vector machine (SVM) for the multi-qubit discriminator. We measure any residual thermal population in the prepared Fock states using an $ef$ Rabi experiment and apply a linear transformation to map the state vector from the prepared basis back to the bare Fock states.
For experiments measuring the two-point correlation function, two single-qubit rotation pulses are applied sequentially at the readout location before the multiplexed readout.
Following the readout, flux balancing pulses are applied to cancel the net current flowing into each flux line during the sequence, thereby preventing unwanted qubit frequency drifts.

The experiments are performed with a typical cycle period of 300 $\mu$s, providing sufficient idle time between shots for the qubits and resonators to relax to their equilibrium ground states. 

\subsubsection{Phase calibration of single qubit rotations for correlation measurements} 

To measure two-qubit correlations after the target evolution, the qubits are rapidly detuned from the lattice configuration, and sequential single-qubit rotations are applied before measurement in the computational basis. During this process, qubit-dependent dynamical phases arise from idling during other rotations and from the finite frequency shift between the lattice and readout configurations. These phases must be calibrated to ensure accurate measurements of $XX$ and $YY$. 
We calibrate these relative dynamical phases by preparing the symmetric highest-energy single-excitation eigenstate of the four-qubit system, which has a known uniform phase profile. For each qubit pair, we sweep the phase of the second rotation pulse and identify the phase that recovers the expected relative phase of the prepared state. The extracted phase offsets are incorporated into subsequent single-qubit rotations used for correlation measurements, ensuring that the measured observables correspond to the intended Pauli operators. 

\subsection{Experimental dataset sizes}

For the population measurements underlying Fig.~\ref{fig:2} and the extracted decay rates in Fig.~\ref{fig:3}, each experiment is typically repeated for $16{,}000$ shots, giving binomial statistical uncertainties below $0.4\%$ for single-qubit populations.

For the decay dynamics data in Fig.~\ref{fig:4}, the extracted decay rates are highly sensitive to fluctuations, particularly at short times and when the population change is small. Each sequence is therefore repeated for $144{,}000$ experimental runs at each time point, obtained from 9 independently acquired datasets, each containing $16{,}000$ single-shot measurements. Readout calibration is performed independently for each dataset to mitigate slow drifts in the measurement chain. Consequently, statistical uncertainties within each dataset are negligible compared to the dataset-to-dataset variations, which dominate the reported uncertainties described below in Sec.~\ref{sec:SI_rate-extraction}.

For the correlation measurements in Fig.~5x, each Pauli observable is measured with $64{,}000$ experimental shots acquired from 4 independent datasets. Both the statistical uncertainties and the standard error of the mean across datasets are small compared to systematic errors arising from single-qubit rotation imperfections. Details of these systematic errors and their mitigation are provided in Sec.~\ref{sec:SI_extradata-correlation}.

\section{Rate extraction and uncertainty estimation}
\label{sec:SI_rate-extraction}

The population $N(t)$ is measured on a nonuniform time grid spanning both short-time superradiant and long-time subradiant dynamics. To balance temporal resolution and noise sensitivity across this range, we extract the decay rates using two complementary procedures:

\begin{enumerate}
    \item {Local linear regression (short-time, $t \lesssim 150\,\mathrm{ns}$).}
    The slope at time $t_i$ is obtained from a least-squares linear fit over all data points $\{(t_j, y_j)\}$ within a symmetric window $W(t)$ of width $80\,\mathrm{ns}$:
    \begin{equation}
    \frac{dy}{dt}\Big|_{t_i}
    =
    \frac{\sum_{j\in W(t)} (t_j-\bar{t})(y_j-\bar{y})}
    {\sum_{j\in W(t)} (t_j-\bar{t})^2},
    \end{equation}
    where $\bar{t}$ and $\bar{y}$ are the averages over the window.

    \item {Three-point finite difference (long-time, $t \gtrsim 150\,\mathrm{ns}$).}
    At longer times, the spacing between adjacent points exceeds the $80\,\mathrm{ns}$ regression window, so the rate at $t_i$ is estimated using $t_i$ and its two nearest neighbors $t_{i\pm1}$. For nonuniform time steps, we use the second-order accurate central-difference formula
    \begin{equation}
    \frac{dy}{dt}\Big|_{t_i}
    =
    \frac{
    (t_i - t_{i-1})^2\, y_{i+1}
    +
    \big[(t_{i+1}-t_i)^2 - (t_i - t_{i-1})^2\big]\, y_i
    -
    (t_{i+1}-t_i)^2\, y_{i-1}
    }{
    (t_i - t_{i-1})(t_{i+1}-t_i)(t_{i+1}-t_{i-1})
    }. 
    \end{equation}
    This is implemented numerically using the \texttt{numpy.gradient} function.

\end{enumerate}

Using the procedures above, two related decay rates are extracted from the measured population $N(t)$:
\begin{itemize}
    \item The emission rate $R(t) = -\,dN/dt$, obtained from $y=N(t)$.
    \item The normalized decay rate $\Gamma(t) = -\frac{1}{N}\frac{dN}{dt} = \frac{d}{dt}\log N(t)$, obtained from $y=\log N(t)$.
\end{itemize}

\paragraph{Uncertainty estimation.}
Uncertainties of decay rates are determined from the variation across the six independent datasets. For each time point, we report the standard error of the mean (SEM),
\begin{equation}
\mathrm{SEM}(t) = \frac{\mathrm{std}\big(\{x_s(t)\}_{s=1}^{N_\text{set}}\big)}{\sqrt{N_\text{set}}},
\end{equation}
where $x_s(t)$ is the extracted rate from dataset $s$, and the number of independent datasets $N_\text{set}$. This captures slow drifts and systematic fluctuations not reflected in intra-dataset statistical noise.

\section{Theoretical modeling}
\label{sec:SI_modeling}

We begin by describing the detailed model used for fully time-dependent simulations of the experiments, corresponding to the theory results shown in Fig.~\ref{fig:4} and the additional numerical results presented in Sec.~\ref{sec:SI_extradata-exp_compare_sim}.
We then discuss two effective models: one for estimating the decay rates of single-excitation states, used to model the data in Fig.~\ref{fig:2}(c) and Fig.~\ref{fig:3}, and another for calculating the many-body decay rates underlying the multi-excitation collective decay dynamics.

\subsection{Full system model}
\label{sec:SI_full-model}

\textbf{Coherent Hamiltonian:}
The total Hamiltonian of the system comprises the transmon qubits coupled to the resonators.
The transmon array forms as an interacting Bose-Hubbard lattice, written in the rotating frame of the unmodulated qubits as:
\begin{equation}
 H_{\text{qubit}}/\hbar = \sum_{i}{J_{i,i+1} (a_i^\dagger a_{i+1} + a_i a_{i+1}^\dagger)} 
 + \sum_{i}{J_{i,i+2} (a_i^\dagger a_{i+2} + a_i a_{i+2}^\dagger)} 
 +\sum_i{\frac{U_i}{2}n_i(n_i-1)}
\end{equation}
where $a^\dagger_i$ ($a_i$) is the creation (annihilation) operator for transmon $i$, with nearest neighbor tunneling $J_{i,i+1}$ and next nearest neighbor tunneling $J_{i,i+2}$, and effective on-site interaction $U_i$ from the anharmonicity of the transmon. 
All experiments in this work use only the lowest two levels of each transmon qubit. 
Hence, the essential physics can be captured as an $XY$ spin chain by replacing $a_i(a^\dagger_i)$ with $\sigma^-_i(\sigma^+_i)$ and having $n_i \in \{0,1\}$.
In our full numerical modeling, we retain three levels per transmon and next-nearest-neighbor tunneling to provide a more accurate comparison with experimental data.

The engineered waveguide coupling of the qubits is enabled via frequency modulation that places the qubit sideband near the readout resonators, thereby enhancing Purcell decay. The modulated qubit sideband frequency $\omega'_\text{q} = \omega_\text{q}+\omega_\text{mod}$ is experimentally set close to the average frequency of all four resonators at roughly $6.25$\,GHz. We model the resonators in the rotating frame of $\omega'_\text{q}$:
\begin{equation}
    H_{\text{res}}/\hbar = \sum^{4}_{i=1} \Delta_\text{eff}^i b^\dagger_i b_i\,, 
\end{equation}
where $b_i$ ($b^\dagger_i$) is the annihilation (creation) operator for readout resonator $i$. For this work, the resonator population remains small during the relatively slow collective decay.

By applying an appropriate qubit-frequency modulation, we generate sideband 
interactions between the lattice and the resonators of the form:
\begin{equation}
    H_{\text{int}}/\hbar = -\sum^{4}_{i=1}g_\text{eff}^i (b^\dagger_i a_i 
    e^{\mathrm{i}\phi_i} + b_i a^\dagger_i e^{-\mathrm{i}\phi_i} )
\end{equation}
where the effective sideband coupling rates with typical values of $2\pi\times (2-8)$\,MHz are chosen to yield the desired engineered single-qubit waveguide 
decay rate of $15 \times 2\pi$ kHz at the average resonator frequency. We can also control the interaction phases $\phi_i$ between each qubit-resonator pair. 
Since the flux modulation is relatively weak, the tunneling $J$ is not significantly modified by the modulation.
\smallskip

\textbf{Lindbladian and master equation:}

The engineered collective decay happens via both right- and left-moving modes of the continuous waveguide (denoted by $s=+1$ and $s=-1$ respectively), as described by the Lindblad collapse operators:
\begin{equation}
c_\text{wg}^{s}
= \sum_{i=1}^4 \sqrt{\frac{\kappa_i}{2}}\,
e^{\,\mathrm{i} s (k_0 d) i}\,
b_i,
\qquad s=\pm 1,
\end{equation}
In addition to the programmable modulation phase $\phi_i$, $k_0 d$ describes the propagation phase between neighboring qubits, set by the coupler position on the waveguide and taken to be frequency-independent given that the emission frequency window is small compared to the emission frequency. Here $\kappa_j$ is the linewidth of the readout resonators.

The estimated $k_0d \approx 10^\circ$ in our device, as discussed above, results in a slight reduction in the interference contrast of the waveguide emission. The contrast is $\cos(k_0 d)\approx 0.985$, a much smaller effect than other effects that also cause imperfect interference, most notably from the frequency dependence of the collective decay due to different resonator detunings and linewidths.
Therefore, we take the approximation that $k_0 d \approx 0$, such that the total waveguide decay, summing over left and right propagating modes, is
\begin{equation}
c_\text{wg} = \sum^4_{i=1} \sqrt{\kappa_i}\,b_i
\end{equation}
The waveguide also mediates coherent interaction between qubits \cite{Lalumiere2013-ck}. Since $d\ll\lambda$ and $g_\text{eff} \ll J$, the waveguide-mediated qubit-qubit coupling is negligible in our experiments compared to the existing tunneling $J$. 

Since we measure approximately the same thermal population in all four resonators $n_\text{th}^\text{R} \approx 2.5$\%, and the waveguide is broadband around the resonator frequency, it is reasonable to assume that the waveguide will have the same effective thermal population at the frequencies of the collective emission. With this finite temperature, the collective decay is represented as a Lindblad superoperator:
\begin{equation}
    \mathcal{L}_{\text{wg}}\rho = (1+n_\text{th}^\text{R})\mathcal{D}[c_\text{wg}]\rho +n_\text{th}^\text{R}\mathcal{D}[c_\text{wg}^\dagger]\rho\,.
\end{equation}
where the standard Lindblad dissipator is $\mathcal{D}[c]\rho = c \rho c^\dagger - \tfrac{1}{2}\{c^\dagger c,\rho\}$, and $\rho$ is the density matrix of the system of qubits and resonators.

\smallskip
\noindent \textbf{Other intrinsic decoherence:}
In addition to the desired collective decay above, we also include other sources of dissipation in the full numerical model: 

\begin{itemize}
    \item \textbf{Intrinsic Purcell-induced collective decay.} This is the Purcell decay $\gamma_\text{purc}$ of the qubit at the unmodulated frequency via the resonator. Since the detuning $\Delta$ is much larger than $J$ here, the rate is largely frequency-independent across the different eigenstates of the coupled qubit array. We model this as a separate decay channel that effectively acts directly on the qubits, where the measured $\gamma_\text{purc} \approx 2\pi\times 2$\,kHz:
    \begin{equation}
        c_\text{purc} = \sum^4_{i=1} \sqrt{\gamma_\text{purc}}\,a_i
    \end{equation}
    From the data in Fig.\,~\ref{fig:int_col_decay}, this intrinsic collective decay channel has an equilibrium thermal population of $n_\text{th}^\text{purc} \approx 5.4\%$ in average. This is included as:
    \begin{equation}
        \mathcal{L}_{\text{purc}}\rho = (1+n_\text{th}^\text{purc})\mathcal{D}[c_\text{purc}]\rho +n_\text{th}^\text{purc}\mathcal{D}[c_\text{purc}^\dagger]\rho
    \end{equation}

    \item \textbf{Independent non-radiative qubit relaxation:} 
    \begin{equation}
        \mathcal{L}_{\text{nr}}\rho = \sum^4_{i=1}(1+n_\text{th}^{Q_i}) \gamma_\text{nr}^i \mathcal{D} [a_i]\rho 
        + \sum^4_{i=1} n_\text{th}^{Q_i}\gamma_\text{nr}^i\mathcal{D} [a^\dagger_i]\rho
    \end{equation}
    
    \item \textbf{Single qubit dephasing:} As discussed above, the eigenstates of the coupled qubit array are, to lowest order, insensitive to local qubit dephasing. Hence, local dephasing is not included in our modeling.

\end{itemize}

\noindent Finally, the full Lindblad master equation reads:
\begin{equation}
    \dot{\rho} = -\mathrm{i}\left[ (H_{\text{qubit}} +H_{\text{res}}+ H_{\text{int}}),\rho\right] + ( \mathcal{L}_{\text{wg}} +  \mathcal{L}_{\text{purc}} + \mathcal{L}_{\text{nr}} )\rho
\end{equation}
which we implement and solve numerically in \texttt{QuTiP}.

\subsection{Decay rates of single-excitation eigenstates}
\label{sec:SI_single_excitation_decay}
The collective decay rates of the single-excitation eigenstates can be obtained analytically within an effective waveguide-coupled model. In the ideal Dicke case, where all emitters couple identically to a common radiation mode and coherent interactions are negligible, the fully symmetric state
\begin{equation}
    \ket{\psi_\mathrm{sym}} 
    = \frac{1}{\sqrt{N_q}} 
    \sum_{i=1}^{N_q} \sigma_i^+ \ket{G}
\end{equation}
radiates at the enhanced rate $\Gamma = N_q \Gamma_0$, where $\ket{G}$ denotes the state with all qubits in the ground state. The remaining $(N_q-1)$ orthogonal single-excitation states are dark in this limit.

In our system, with strong nearest-neighbor exchange interactions $J$, the local excitations are hybridized. The single-excitation eigenstates can be written generally as
\begin{equation}
    \ket{m} 
    = \sum_{i=1}^{N_q} \psi_m(i)\, \sigma_i^+ \ket{G}, \quad 
    m=1,\dots,N_q
\end{equation}
with eigenenergies $E_m$ in the rotating frame of the qubits. For a degenerate $N_q$-site lattice with open boundary conditions, the normalized modes are standing waves with wavefunctions and the corresponding eigenenergies
\begin{equation}
    \psi_m(i) =
    \sqrt{\frac{2}{N_q+1}}
    \sin\!\left( \frac{\pi m i}{N_q+1} \right), \quad
    E_m = 2J \cos\!\left( \frac{\pi m}{N_q+1} \right).
\end{equation}

\textit{\textbf{Frequency-independent emission model:}}
We first consider an effective description in which the qubit–waveguide coupling is assumed to be frequency independent. For a normalized single-excitation eigenstate
\begin{equation}
    \ket{m} = \sum_{i=1}^{N_q} \psi_m(i)\, \sigma_i^+ \ket{G},
\end{equation}
The total emission rate into the waveguide (including both left- and right-propagating modes) is
\begin{equation}
    \Gamma_m = |A_l(m)|^2 + |A_r(m)|^2,
\end{equation}
with emission amplitudes
\begin{equation}
    A_{r,l}(m)
    =
    \sum_{i=1}^{N_q}
    \sqrt{\gamma_i}\,
    e^{\pm \mathrm{i} k_0 x_i + \mathrm{i}  \phi_i}\,
    \psi_m(i).
\end{equation}
Here $\gamma_i$ is the effective single-qubit decay rate into the waveguide, $x_i$ is the qubit position, and $\phi_i$ is the programmable phase set by the flux modulation.

\textit{\textbf{Frequency-dependent resonator response:}}
The previous model assumes frequency-independent coupling to the waveguide. In reality, the qubit–waveguide interaction is mediated by finite-bandwidth resonators. Because the eigenstates are split in energy by $J$, each mode $\ket{m}$ emits at a distinct frequency set by $E_m$, and the decay rate must therefore include the resonator spectral response.

For a resonator coupled to qubit $i$ with linewidth $\kappa_i$ and effective detuning $\Delta_{\mathrm{eff},i}$, the effective emission rate takes the Lorentzian form
\begin{equation}
    \gamma_i(E_m)
    =
    \frac{g_{\mathrm{eff},i}^2\,\kappa_j}
         {(E_m - \Delta_{\mathrm{eff},i})^2 + (\kappa_i/2)^2},
\end{equation}
where $g_i$ is the modulation-induced coupling strength. In the experiment, the flux modulation amplitudes are chosen such that the single-qubit decay rates at the center of the band are uniform across all qubits,
\begin{equation}
    \gamma_{\mathrm{wg}}
    =
    \gamma_i(0)
    =
    \frac{g_{\mathrm{eff},i}^2\,\kappa_i}
         {\Delta_{\mathrm{eff},i}^2 + (\kappa_i/2)^2}.
\end{equation}
Since this calibration frequency ($E=0$) satisfies $|\Delta_{\mathrm{eff},i}| \gg \kappa_i$, the decay is in the dispersive limit with
\begin{equation}
    \gamma_{\mathrm{wg}}
    \approx
    \frac{g_{\mathrm{eff},i}^2}{\Delta_{\mathrm{eff},i}^2}\,\kappa_i.
\end{equation}

The full frequency-dependent collective decay rate is therefore
\begin{equation}
    \Gamma_m
    =
    \left|
    \sum_{i=1}^{N_q}
    \sqrt{\gamma_i(E_m)}\,
    e^{+ \mathrm{i}  k_0 x_i + \mathrm{i} \phi_i}\,
    \psi_m(i)
    \right|^2
    +
    \left|
    \sum_{i=1}^{N_q}
    \sqrt{\gamma_iE_m)}\,
    e^{- \mathrm{i}  k_0 x_j + \mathrm{i}  \phi_i}\,
    \psi_m(i)
    \right|^2.
    \label{eq:SI_freq-dep-decay}
\end{equation}

In the non-interacting limit ($J=0$), all single-excitation states are degenerate and emit at the same frequency, so $\gamma_i(E_m)$ is identical for all $m$ and spectral filtering plays no role. For $J \neq 0$, the interaction-induced energy splitting produces mode-dependent decay rates.

When the level spacing satisfies $J/N_q \gg \gamma_i(E_m)$, emission from different eigenstates becomes spectrally distinguishable. In that regime, photons emitted from different eigenmodes do not interfere. While this does not affect the decay of individual eigenstates, it becomes relevant when analyzing the decay of multi-excitation initial states.

\begin{figure}[!htbp]
    \centering
    \includegraphics[width=0.95\columnwidth]{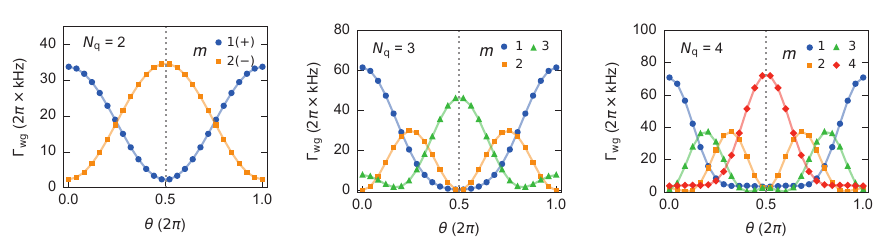}
    \caption{
    Simulated collective decay rates $\Gamma_m$ of single-excitation eigenstates versus $\theta$ for lattices of different sizes, modeled using the frequency-dependent resonator response. Experimental measurements are shown in Fig.\,\ref{fig:phase_vector_exp}.}
    \label{fig:phase_vector_simulation} 
\end{figure}

\begin{figure}[!htbp]
    \centering
    \includegraphics[width=0.9\columnwidth]{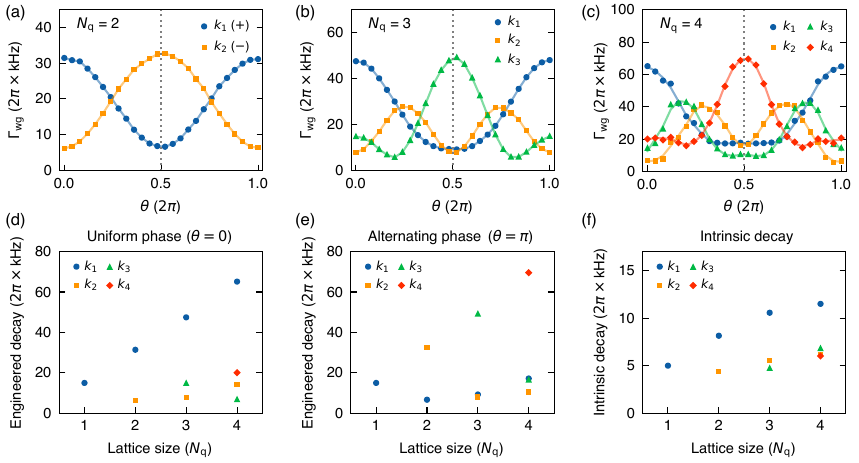}
    \caption[]{Measured decay rates of single-excitation eigenstates. Top row: Decay rates versus $\theta$ for different lattice sizes. Bottom row: Lattice-size scaling of the eigenstate decay rates for $\theta=0$ (Fig.~\ref{fig:2}c), $\theta=\pi$, and the intrinsic decay without engineered waveguide coupling.}
    \label{fig:phase_vector_exp} 
\end{figure}

For the decay of single-excitation eigenstates (all experiments in Fig.~\ref{fig:2} and Fig.~\ref{fig:3}), we choose the sideband frequency to match the average frequency of the two middle resonators $R_1$ and $R_2$, yielding effective resonator detunings of $\Delta_{\mathrm{eff}}^i = \omega_{\mathrm{r}} - \omega_{\mathrm{q}} - \omega_{\mathrm{mod}} \approx \{68.5, 23.3, -23.5, -74.3\} \times 2\pi$~MHz. We then calibrate the engineered single-qubit collective decay rate to $\gamma_{\mathrm{wg}} = 15 \times 2\pi$~kHz. In Fig.~\ref{fig:phase_vector_simulation}, we calculate the decay rates of the single-excitation eigenstates $\Gamma_m$ using Eqn.\,\ref{eq:SI_freq-dep-decay}. The results are shown for lattice sizes $N_q=2,3,4$ as a function of the waveguide coupling phase difference between neighboring qubits, $\theta = \phi_{i+1}-\phi_i$.

The corresponding experimental results are shown in the top row of Fig.\,\ref{fig:phase_vector_exp} for $N_q=2$ (using $Q_1$-$Q_2$, corresponding to the data in Fig.~\ref{fig:2}c), $N_q=3$ (using $Q_0$-$Q_1$-$Q_2$), and $N_q=4$ (using all four qubits). We observe good agreement with the effective model. The dominant discrepancies are attributed to residual flux crosstalk, which produces inhomogeneous decay rates and distorts the modulation amplitudes, leading to weakly time-dependent engineered decay rates.

By extracting the decay rates at fixed $\theta$ for different $N_q$, we obtain the lattice-size scaling of the engineered decay rates shown in Fig.~\ref{fig:3} for uniform-phase coupling ($\theta=0$), reproduced in Fig.\,\ref{fig:phase_vector_exp}(d). In (e), we show the scaling for alternating-phase coupling ($\theta=\pi$), where the superradiant states correspond to $m=N_q$, the lowest-energy eigenstate with alternating phases on neighboring sites. Clear linear scaling with lattice size is again observed.
In (f), we show the lattice-size dependence of the measured intrinsic qubit decay without flux modulation. The similar scaling behavior, albeit at a much lower rate, indicates that the intrinsic decay also contains a collective contribution arising from Purcell decay through the resonators into the common waveguide.

\subsection{Many-body decay rate via projection-operator formalism}
\label{sec:SI_manybody-rates}
\noindent \textit{\textbf{Projection-operator formalism:}} To compute decay rates in the presence of non-uniform resonator detunings $\{\Delta_i\}$ and linewidths $\{\kappa_i\}$ beyond the single-excitation manifold, we employ a projection-operator formalism to calculate the effective decay rates of the lattice. By adiabatically eliminating the fast resonator degrees of freedom, we derive an effective non-Hermitian Hamiltonian acting within the lattice (qubit) manifold, allowing us to extract transition-resolved decay rates. 

We divide the Hilbert space into two orthogonal sectors. The $\mathcal{P}$ subspace consists of lattice eigenstates with all resonators in their ground state, while the $\mathcal{Q}$ subspace contains states with one photon in any of the four resonators. Because the resonators decay into a common waveguide with rates $\kappa_i$, dynamics within $\mathcal{Q}$ occur on a rapid timescale of $\sim 1/\kappa_i$ and can be adiabatically eliminated.

Treating the sideband interaction as a perturbation, we first write the non-Hermitian Hamiltonian of the free qubit lattice and resonators as
\begin{equation*}
H = H_{\mathrm{qubit}} + H_{\mathrm{res}} - \frac{\mathrm{i}}{2} \hbar c^\dagger c, 
\quad \text{with} \quad c = \sum_{i=1}^{4} \sqrt{\kappa_i}\, b_i.
\end{equation*}

Expanding the dissipative term gives
\begin{equation*}
c^\dagger c = \sum_{i,j} \sqrt{\kappa_i \kappa_j}\, b_i^\dagger b_j\,.
\end{equation*}

The cross terms ($i\neq j$) arise because all resonators couple to the same waveguide continuum, generating correlated emission and thus dissipative coupling between resonators.

In the single-resonator-excitation basis $\{|1_1\rangle,\dots,|1_4\rangle\}$, the Hamiltonian involving the excited subspace $\mathcal{Q}$ is:
\begin{equation*}
(QHQ)_{ij}/\hbar=
\Delta^i_{\mathrm{eff}} \delta_{ij} -
\frac{\mathrm{i} }{2}\sqrt{\kappa_i\kappa_j}.
\end{equation*}

Following the projection-operator formalism, the effective Hamiltonian acting within $\mathcal{P}$ takes the form:
\begin{equation*}
H_{\mathrm{eff}}(z)=PHP+
PHQ \frac{1}{z - QHQ} QHP,
\end{equation*}
where $z$ is evaluated at the transition frequency.

The lattice-resonator interaction
\begin{equation*}
H_{\mathrm{int}}/\hbar=-\sum_{i=1}^{4}
g^i_{\mathrm{eff}} b_i^\dagger \sigma_i^- e^{\mathrm{i}\phi_i}
+
\mathrm{h.c.}
\end{equation*}
gives transition matrix elements from the initial state $|\Psi_{\mathrm{i}}\rangle$ to the final state $|\Psi_{\mathrm{f}}\rangle$ through the sideband interaction on the site $i$:
\begin{equation*}
V_{\mathrm{i}\to \mathrm{f}, i}=
-g^i_{\mathrm{eff}} e^{\mathrm{i} \phi_i}
\langle \Psi_\mathrm{f}|\sigma_j^-|\Psi_\mathrm{i}\rangle.
\end{equation*}

Defining the vector $\mathbf{V}_{\mathrm{i}\to \mathrm{f}}$, we express the complex self-energy $\Sigma_{\mathrm{i}\to \mathrm{f}}$ corresponding to the transition frequency $\omega_{\mathrm{if}}$ as:
\begin{equation*}
\Sigma_{\mathrm{i}\to \mathrm{f}}(\omega_{\mathrm{if}})=\mathbf{V}_{\mathrm{i}\to \mathrm{f}}^\dagger
\left[
\omega_{\mathrm{if}} I - QHQ
\right]^{-1}
\mathbf{V}_{\mathrm{i}\to \mathrm{f}}.
\end{equation*}

The decay rate is obtained from the imaginary part, which corresponds to irreversible emission into the waveguide:
\begin{equation*}
\Gamma_{\mathrm{i}\to \mathrm{f}}
=
-2\,\mathrm{Im}\,\Sigma_{\mathrm{i}\to \mathrm{f}}(\omega_{\mathrm{if}}),
\end{equation*}
while the real part produces a Lamb shift that arises from virtual emission into detuned resonator modes followed by reabsorption:
\begin{equation*}
\delta\omega_{\mathrm{i}\to \mathrm{f}}=
\,\mathrm{Re}\,\Sigma_{\mathrm{i}\to \mathrm{f}}(\omega_{\mathrm{if}}).
\end{equation*}

\begin{figure*}[!htbp]
    \centering
    \includegraphics[width=0.85\columnwidth]{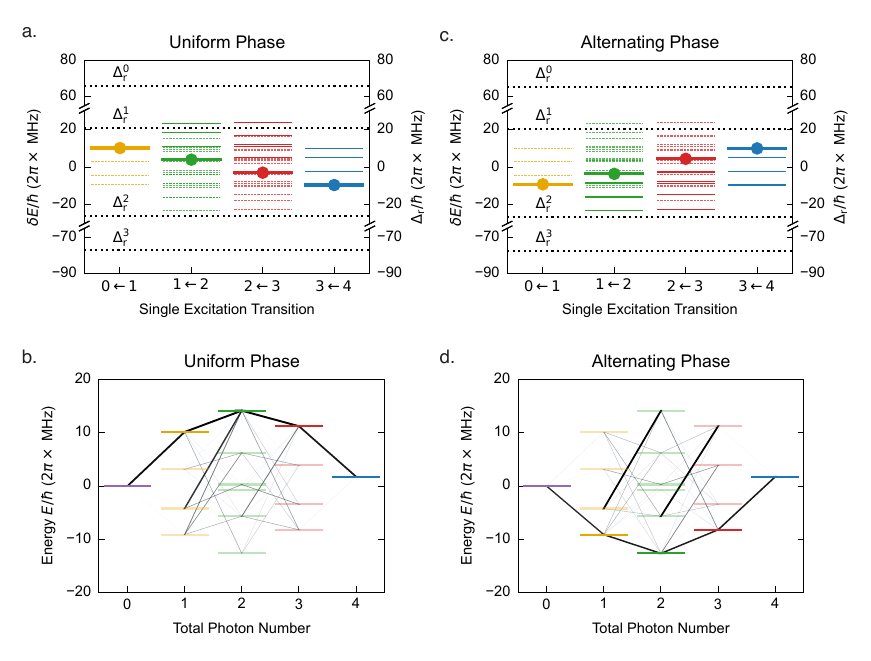}
    \caption[]{Simulated emission Spectra and engineered emission rates. (a-b) uniform-phase $\theta = 0$; (c-d) alternating-phase $\theta = \pi$.
    }
    \label{fig:transition_decay_combined} 
\end{figure*}

\smallskip
\textit{\textbf{Emission spectra and engineered emission rates:}}

We now apply the projection-operator formalism to calculate the emission spectra and many-body emission rates for the four-qubit experiments in Fig.~\ref{fig:4} and Fig.~\ref{fig:5}, shown in Fig.~\ref{fig:transition_decay_combined}. These plots help show how the engineered dissipation couples to different many-body transitions under either uniform-phase or alternating-phase collective decay ($\theta=0$ and $\theta=\pi$).

Figure~\ref{fig:transition_decay_combined}(a,c) shows the energy of all transitions connecting neighboring excitation number manifolds ($N\to N-1$). These transitions can be grouped into three categories: The thick solid lines with round markers correspond to transitions within the superradiant manifold and form the dominant collective decay channel. The thin solid lines represent transitions from the superradiant manifold into subradiant states. Finally, the dotted lines denote transitions entirely within the subradiant manifold, which are only weakly coupled to the collective dissipation and contribute little to the observed superradiant emission.

The horizontal black dotted lines indicate the effective resonator detunings $\Delta_{\mathrm{eff}}^i$ relative to the collective emission frequencies. If a resonator detuning is near-resonant to one of the transitions within the resonator linewidth $\kappa_r$, the corresponding resonator can induce unwanted local dissipation that competes with the engineered collective decay. In the present device, the relatively small spacing between resonator frequencies (for example, $\sim2\pi\times45$~MHz between $R_1$ and $R_2$), together with the fact that most transitions lie within an energy band of approximately $\pm4J\sim2\pi\times50$~MHz, requires careful choice of the effective detunings in order to suppress unwanted uncorrelated decay processes that would otherwise mask the collective decay dynamics.

For the uniform-phase configuration, offsetting the sideband frequency by approximately $2\pi\times3$\,MHz from the midpoint between $R_1$ and $R_2$ yields the effective detunings
\[
\Delta_{\mathrm{eff}}^i \approx 2\pi\times\{66.3,\,20.6,\,-26.2,\,-76.7\}\,\mathrm{MHz}.
\]
For the alternating-phase configuration, an additional $2\pi\times(-0.5)$\,MHz detuning is applied, resulting in $\Delta_{\mathrm{eff}}^i \approx 2\pi\times\{65.8,\,20.1,\,-26.7,\,-77.2\}\,\mathrm{MHz}$ . These effective detunings are shown in Fig.~\ref{fig:transition_decay_combined}(a,c) and are used in both the experiment and numerical simulations.

We note that for decay of the single-excitation eigenstates ($N=1\rightarrow0$), all transition frequencies lie within approximately $\pm2J\approx \pm2\pi\times12$\,MHz of $(\omega_{\mathrm{q}}+\omega_{\mathrm{mod}})$. As a result, when the sideband frequency is chosen near the midpoint between $R_1$ and $R_2$, these transitions remain at least $\sim2\pi\times10$\,MHz detuned from the resonators. 
Consequently, for the experiments in Fig.~\ref{fig:2} and Fig.~\ref{fig:3} (Sec.~\ref{sec:SI_single_excitation_decay}), we choose the sideband frequency near the midpoint between $R_1$ and $R_2$, corresponding to effective detunings
$\Delta_{\mathrm{eff}}^i \approx 2\pi\times\{68.5,\,23.3,\,-23.5,\,-74.3\}\,\mathrm{MHz}$,
without having to consider significant contributions from the unwanted transitions considered here.

Figures~\ref{fig:transition_decay_combined}(b,d) show the corresponding engineered decay rates $\Gamma_{i\rightarrow j}$ between many-body eigenstates, where the thickness of each transition is proportional to the calculated collective emission rate. Under uniform-phase coupling, a dominant high-rate decay pathway emerges along the highest-energy states of each excitation manifold, corresponding to the primary Dicke-like superradiant ladder. In contrast, transitions into and within the subradiant manifold remain suppressed. The large contrast in transition rates demonstrates that the collective superradiant decay dominates the dynamics despite the presence of residual leakage channels. Figure~\ref{fig:transition_decay_combined}(d) shows the corresponding many-body transition rates for alternating-phase coupling, where the engineered phase pattern qualitatively restructures the collective emission pathways such that the dominant superradiant decay now proceeds through the lowest-energy eigenstates.

\subsection{Early time dynamics}
\label{sec:SI_earlytime}

When the sideband modulation is turned on instantaneously, we observe strong oscillations in both the emission rate $R(t)$ and the normalized decay rate $\Gamma(t)$. This occurs because the initial coherent preparation generates Dicke-like states that are eigenstates of the bare lattice Hamiltonian. Quenching this system into the interacting lattice-resonator regime abruptly projects these states into superpositions of the combined system's dressed states, driving a rapid back-and-forth exchange of photons between the qubits and resonators.

\begin{figure}[!htbp]
    \centering
    \includegraphics[width=0.55\columnwidth]{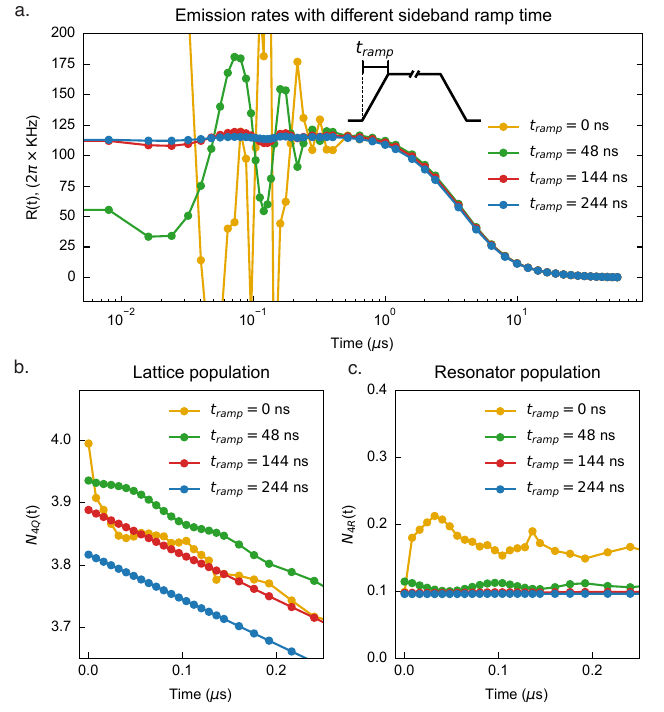}
    \caption[]{Early time decay dynamics from fully inverted initial state in $N_q=4$, with different sideband ramp durations. }
    \label{fig:SI_difframptime} 
\end{figure}

To mitigate this, we turn on the sideband interaction adiabatically by applying a linear ramp to the modulation. The interaction Hamiltonian $H_{\text{Int}}$ thus becomes time-dependent:
\begin{equation}
    H_{\text{Int}}(t)/\hbar = -\lambda(t)\sum^{4}_{i=1}g_i \left(b^\dagger_i \sigma^-_i e^{\mathrm{i}\phi_i} + b_i \sigma^+_i e^{-\mathrm{i}\phi_i} \right)\,, 
\end{equation}
where the envelope function $\lambda(t)$ takes the form
\begin{equation}
\lambda(t) = \begin{cases}
    t/t_{\text{ramp}}, & t < t_{\text{ramp}}\,,\\
    1, & t \geq t_{\text{ramp}}\,.
\end{cases}
\end{equation}
At the end of the collective decay, the sideband interaction is ramped down with the same duration. The simulated decay dynamics from a fully inverted state, shown in Fig.\,\ref{fig:SI_difframptime}, highlight the effect of varying $t_{\text{ramp}}$. For short ramps, abrupt photon transfer into the resonators leads to oscillations in the emission rate. We use $t_{\text{ramp}} = 244$ ns in the experiments to suppress these early time oscillations.

\subsection{Effect of thermal population}

\paragraph{Qubit thermal population.}
In most experiments, we use a heralding measurement to remove thermal population in the multi-qubit initial state. During the collective decay experiments, the qubit relaxation is dominated by the engineered decay rate $\gamma_\text{wg}$ into the waveguide and the intrinsic Purcell decay through the resonators. Other decay channels that act locally on each qubit are small in comparison. As a result, any effective ``re-thermalization'' of the qubits primarily occurs through the resonators and the waveguide environment.
At long times, the qubits relax to an effective thermal population determined by the temperature of the waveguide mode to which the qubits couple via the resonators. Because the coupling to the waveguide occurs through the collective decay channel, this thermal population preferentially occupies the symmetric eigenmode of the qubit array. This behavior is evident in Fig.~\ref{fig:int_col_decay}, where the long-time population approaches a higher value only for the symmetric eigenstate.

\paragraph{Resonator thermal population.}
A pre-existing thermal photon population in the resonator modifies the qubit relaxation rates via stimulated processes. At early times, immediately after the engineered qubit-waveguide coupling is enabled, the dominant effect on the decay of an initially excited qubit is a small enhancement of the relaxation rate by a factor $(1+n_\text{th}^r)$.
In practice, this leads to three observable effects in the decay curve $N(t)$: a slightly larger initial value of $|dN/dt|$, a slight increase in the overall relaxation rate, and a long-time population that approaches a nonzero thermal steady state.

\paragraph{Effect on two-qubit correlations.}
If the thermal fluctuations acting on each qubit are independent, the steady state of the system is a product state and the thermal fluctuations have no contributions to any of the connected two-qubit correlators.
In contrast, when the thermal population is introduced through a \emph{collective} decay channel (as in our system, where all qubits couple to the same waveguide mode through the resonators), the bath couples through the collective operators.
As an example, if the steady qubit array state reaches thermal equilibrium with the waveguide, which has $P_\text{th} \sim 2.5\%$ probability to be in the symmetric single-excitation state, and negligible probability in all other non-vacuum states. Because this symmetric state contains coherent superpositions of excitations across different qubits, correlated thermal populations produce small but finite transverse correlations between qubits even in the absence of coherent interactions. In this example, we can estimate:
$ C^{xx/yy}_{i,j} \approx 2 P_\text{th}/N_q \approx 0.01$, and
$ C^{zz}_{i,j} = - (2P_\text{th}/N_q)^2\approx - 1.5\times10^{-4}$.
These estimates are in good agreement with results from the full model.

\section{Additional data and analysis}
\label{sec:SI_additional-data}

\subsection{Collective decay with mismatched emission frequency}
\label{sec:SI_extradata-sideband-detuned}

Here we present additional analysis and modeling of the experiment shown in Fig.\,\ref{fig:2}d, which measures the decay of single-excitation states in two qubits as a function of the detuning $\delta$ and relative phase $\theta$ between the two sideband modulations. The reduction in contrast between $\Gamma_+$ and $\Gamma_-$ as $|\delta|$ increases can be interpreted as arising from the reduced interference between spectrally distinguishable emissions from the two qubits.

Collective emission is governed by interference between two emission amplitudes, described by a collective jump operator
$L=\sigma_1^-+\sigma_2^-$.
We focus on the case where the waveguide coupling has the same initial phase ($\theta=0$) and a frequency-independent decay rate $\gamma_\text{wg}$.
The rate of population change equals the emitted intensity and follows directly from the Lindblad master equation as:
\begin{align}
\frac{d}{dt}N(t)
&=-\gamma_\text{wg}\langle L^\dagger L\rangle \nonumber\\
&=-\gamma_\text{wg}\langle \hat n_1 + \hat n_2
+\sigma_1^+\sigma_2^- +\sigma_2^+\sigma_1^- \rangle \nonumber\\
&=-\gamma_\text{wg}\left[N(t)+2\,\mathrm{Re}\,C_{12}(t)\right],
\label{eq:dNdt_final}
\end{align}
where the two-qubit coherence $C_{12}(t)=\langle\sigma_1^+(t)\sigma_2^-(t)\rangle$ fully determines the interference contribution to the emitted field.
It is useful to define an instantaneous decay rate
\begin{equation}
\Gamma_{\rm inst}(t)
= -\frac{d}{dt}\ln N(t)
= \gamma_\text{wg}\left[1+2\,\frac{\mathrm{Re}\,C_{12}(t)}{N(t)}\right].
\label{eq:Gamma_inst_final}
\end{equation}

\vspace{4pt}
\noindent
{\bf Evolution of $C_{12}(t)$ with detuning:}
For an initial single-excitation state
\begin{equation}
|\psi(\theta_0)\rangle=\frac{|eg\rangle+e^{\mathrm{i} \theta_0}|ge\rangle}{\sqrt2},
\end{equation}
a detuning $\delta$ generates an additional dynamical phase difference,
\begin{equation}
\theta(t)=\theta_0+\delta t.
\end{equation}
Note that, for simplicity, we provide an effective description in which the relative phase and detuning are written directly in the two-qubit state, whereas in the experiments in Fig.\,\ref{fig:2} they are implemented via flux modulation.
As a result of the detuning, the two-qubit coherence oscillates in time with a decaying envelope determined by the total single-qubit dephasing rate:
\begin{equation}
C_{12}(t)\propto e^{\mathrm{i} (\theta_0+\delta t)}e^{-\gamma_\text{deph}t}.
\end{equation}
For two qubits with identical radiative decay $\gamma_\text{wg}$ and pure dephasing $\gamma_\phi$, a useful estimate of the \textit{relative} dephasing rate is
\begin{equation}
\gamma_\text{deph} \simeq \gamma_\text{wg} + 2\gamma_\phi,
\end{equation}
assuming uncorrelated noise on the two qubits.

Two important consequences follow. First, when the emissions are detuned, the decay rate becomes time-dependent: a prepared bright state evolves into a darker state and back as the relative phase evolves. Second, this motivates the analysis of decay over a finite observation window. As we show below, ``spectral distinguishability'' in this experiment is not a static property of the emitters but an operational quantity that depends on the measurement bandwidth through the energy--time uncertainty relation.

\vspace{4pt}
\noindent
{\bf Qualitative behavior in limiting cases:}
To gain an analytical understanding of the detuning dependence of the contrast, we consider the two-qubit coherence averaged over a time window of length $T$:
\begin{equation}
Z(\delta,T)=\int_0^T dt\,C_{12}(t).
\end{equation}
The magnitude $|Z(\delta,T)|$ measures the time-averaged interference contrast, while the phase $\arg Z(\delta,T)$ represents the dynamical phase accumulated over the observation window and determines which initial state exhibits maximal or minimal window-averaged decay. The rectangular window used here differs from the exact weighting implied by the exponential fit, but captures the relevant behavior up to an overall constant factor of order unity.

Substituting the form of $C_{12}(t)$ yields
\begin{equation}
Z(\delta,T)\propto e^{i\theta_0}
\frac{1-e^{-(\gamma_\text{deph}-\mathrm{i} \delta)T}}{\gamma_\text{deph}-\mathrm{i} \delta}.
\end{equation}
which captures the detuning dependence of the observed contrast and phase.

(1) In the short-window limit $T\ll 1/\gamma_\text{deph}$,
\begin{equation}
|Z(\delta,T)|
\approx
\left|\int_0^T dt\,e^{\mathrm{i} \delta t}\right|
=
\left|T\,\mathrm{sinc}\!\left(\frac{\delta T}{2}\right)\right|.
\end{equation}
The characteristic detuning scale is therefore set by the width of the sinc function. If we take the locations of the first zeros of the sinc at $\delta\,T/2=\pm\pi$, that gives the full width of the detuning tolerance window $\Delta\delta \approx 4\pi/T$.
Meanwhile, the phase arg$(Z)$ advances as $\delta t$, hence averaged over the window $T$ has a linear slope $\frac{d}{d\delta} \text{arg}(Z) \approx T/2$. Together, the overall phase shift observed over the detuning window with finite contrast is $(4\pi/T)\times(T/2) \sim 2\pi$. This qualitative estimate is consistent with the measurement and simulation results shown below.

(2) In the opposite limit $T\gg 1/\gamma_\text{deph}$,
\begin{equation}
|Z(\delta,T)|\approx\left|\frac{1}{\gamma_\text{deph}-\mathrm{i} \delta}\right|= \frac{1}{\sqrt{\gamma_\text{deph}^2+\delta^2}},
\end{equation}
so the detuning tolerance no longer decreases with $T$ but instead saturates at
\begin{equation}
\Delta\delta \sim \gamma_\text{deph}.
\end{equation}
Thus, spectral distinguishability in this experiment is determined by the smaller of the measurement bandwidth (set by the analysis window $T$, which reflects the time-energy uncertainty principle) and the intrinsic decoherence rate (specifically, the relative dephasing rate between the two qubits).

\begin{figure}[H]
    \centering
    \includegraphics[width=0.95\columnwidth]{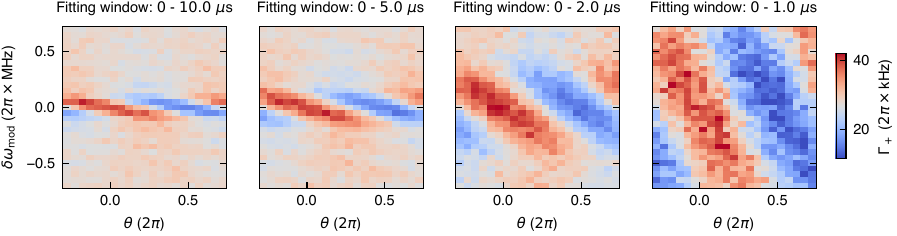}
    \caption[]{Effective decay rate of the two qubit $\ket{+}$ state under mismatched flux modulation frequencies, analyzed using different decay time windows. The higher decay rates compared to the simulation in Fig.\,\ref{fig:SI_detuned-mod-sim} come from intrinsic decay and frequency dependence of the waveguide decay which are not included in the theoretical model.}
    \label{fig:SI_detuned-mod-data} 
\end{figure}

\begin{figure}[H]
    \centering
    \includegraphics[width=0.95\columnwidth]{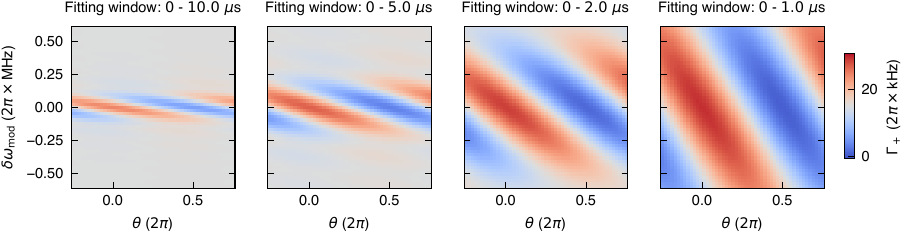}
    \caption[]{Modeling of the experimental results in Fig.\,\ref{fig:SI_detuned-mod-data}.}
    \label{fig:SI_detuned-mod-sim} 
\end{figure}

\vspace{4pt}
\noindent
{\bf Data and numerical modeling:}
In Fig.\,\ref{fig:SI_detuned-mod-data}, we plot the measured decay rate of the symmetric superposition $\psi_+$ as a function of the modulation detuning and initial phase difference of the modulation, when analyzing a finite window starting from $t=0$ for durations of $T = 1,2,5,10\,\mu$s. For each value of $T$, the decay rate is determined by fitting a single exponential to the population dynamics within the finite window.  In Fig.\,\ref{fig:SI_detuned-mod-sim}, we simulate the measured data using the instantaneous decay rate $\Gamma_\text{inst}(t)$ and then apply the same fitting procedure. In the simulation we use $\gamma_\text{wg} = 15$\,kHz and $\gamma_\phi = 0.5$\,kHz. The latter value is chosen to reflect the suppression of local dephasing expected in the presence of large coherent coupling $J$. The simulation shows excellent quantitative agreement with the experimental data.
For larger dephasing, the overall contrast rapidly decreases once the observation window exceeds the dephasing time.

\subsection{Role of dephasing: Example with two qubits}
\label{sec:SI_extradata-qubits-detuned}

For resonantly coupled qubit arrays, we have shown that local dephasing is effectively suppressed by the strong coherent interaction $J$ to stabilize collective decay.

Here we show an experiment that effectively eliminates $J$ in the two-qubit case, thereby restoring the effect of dephasing. We start with a large detuning $\Delta \gg J$ between the two qubits. In this large detuning limit, the single excitation eigenstates are approximately the same as the bare qubit states and effectively uncoupled. The intrinsic dissipation and Purcell decay from the two qubits, in the absence of the flux modulation, are at different frequencies and remain independent.
We then turn on the flux modulation at frequencies that differ by $\Delta$, so that the sideband decay frequencies coincide for the two qubits, enabling interference in the waveguide emission. We tune the modulation amplitudes so that the engineered Purcell decay rate $\Gamma_\text{eff}$ is the same for the two qubits. We prepare the symmetric and antisymmetric superpositions $|\pm\rangle=(|g_1e_2\rangle\pm|e_1g_2\rangle)/\sqrt{2}$ and measure their relaxation, here with effective alternating-phase couplings $\theta=\pi$.

\begin{figure}[H]
    \centering
    \includegraphics[width=0.5\columnwidth]{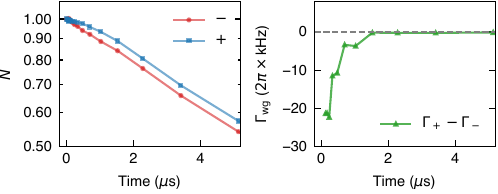}
    \caption[]{Collective decay in the presence of large local dephasing.}
    \label{fig:SI_detuned-2Q} 
\end{figure}

With dephasing $\Gamma_\phi = 1/T_2^* \approx 75$\,kHz exceeding the engineered decay rate, different initial superposition states will dephase toward the same incoherent mixed state during the decay dynamics. This is seen in Fig.~\ref{fig:SI_detuned-2Q} both in population dynamics, and by comparing the instantaneous decay rates for the $\ket{\pm}$ states. The decay rates $\Gamma_\pm$ converge to the same value at a time scale of $T_2^*/2\approx 1\,\mu$s, where the factor of 2 comes from considering the relative dephasing rate between the two qubits.

\subsection{Decay from fully inverted state - Modeling}
\label{sec:decay_from_FIS_model}

The numerical results in Fig.~\ref{fig:4} are obtained from time-dependent simulations of the full model described in Sec.~\ref{sec:SI_full-model}, including the 244\,ns ramp-up and ramp-down of the engineered decay. The simulations use the ideal fully inverted initial state ($N_0=4$), and the qubit populations are sampled at the same time points as in the experiment. From these populations, we calculate the emission rate $R(t)$, normalized decay rate $\Gamma(t)$, and number fluctuations using the same analysis procedure as for the experimental data.

The parameters used in the simulations are as follows: The coherent qubit Hamiltonian $H_{\mathrm{qubit}}$ uses parameters $J_{i,i+1}$ and $J_{i,i+2}$ detailed in Table~\ref{tab:device_parameters}. The effective resonator detunings, $\Delta_{\mathrm{eff}}^i = \omega_{\mathrm{r}} - \omega_{\mathrm{q}} - \omega_{\mathrm{mod}} \approx \{66.3, 20.6, -26.2, -76.7\} \times 2\pi$~MHz, and the sideband coupling rates, $\{g_{\mathrm{eff}}^i\} \approx \{6.53, 2.28, -2.53, -7.93\} \times 2\pi$~MHz.
Furthermore, the Lindblad model includes resonator decay with linewidths $\kappa_{i}$ (see Table~\ref{tab:device_parameters}) with a thermal population of $n^{\mathrm{R}}_{\mathrm{th}} \approx 2.5\%$. Qubit relaxation is also included, comprising independent non-radiative relaxation ($\Gamma_{\mathrm{nr}}\approx 4\times 2\pi$~kHz) with a thermal population of $n^{\mathrm{Q}}_{\mathrm{th}} \approx 0.05$, as well as experimentally measured intrinsic Purcell-induced collective decay ($\gamma_{\mathrm{purc}} \approx 10\times 2\pi$~kHz) with a thermal population of $n^{\mathrm{purc}}_{\mathrm{th}} \approx 5.4\%$.

\subsection{Decay from Dicke states - Modeling}
\label{sec:SI_extradata-exp_compare_sim}

\begin{figure}[!htbp]
    \centering
    \includegraphics[width=0.7\columnwidth]{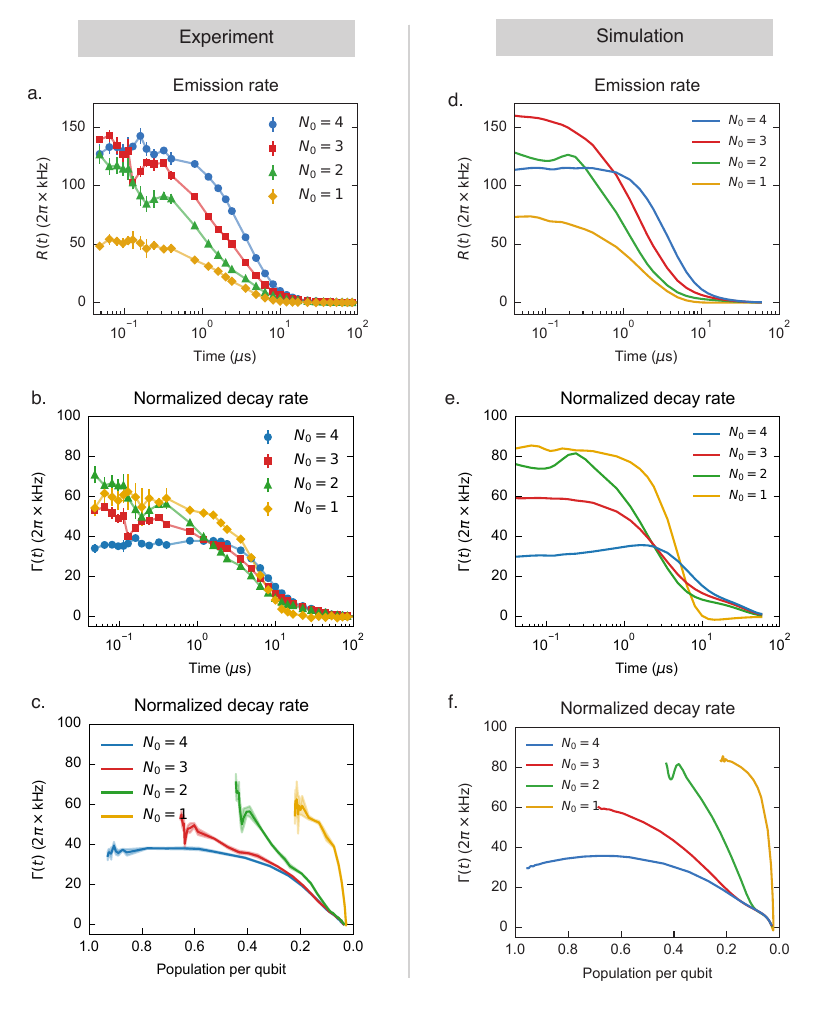}
    \caption[]{Uniform-phase collective decay from different Dicke states: (a)-(c) Experimental data, (a,c) are replotted from Fig.\,\ref{fig:4}(g,h). (d)-(f) Simulations resuts.}
    \label{fig:decay_rate_compare_to_sim} 
\end{figure}

In Fig.~\ref{fig:decay_rate_compare_to_sim}, we present numerical simulations corresponding to the data in Fig.~\ref{fig:4}(g,h) for decay from Dicke-like states. The simulations use ideal highest-energy eigenstates with initial excitation numbers $N_0=1$ to $4$, using the same parameters as in Sec.~\ref{sec:decay_from_FIS_model}.

Qualitatively, the experimental data show a smaller separation in the early-time decay rate among different $N_0$ states than the numerical simulations. We attribute this discrepancy primarily to imperfect state preparation. Although our coherent preparation scheme efficiently prepares the fully inverted state, it introduces larger coherent errors for the partially filled states with $N_0=1,2,3$. The resulting admixture of other excitation-number manifolds in the initial state reduces the contrast between their decay rates.
In addition to state-preparation errors, we observe higher decay rates for nominally subradiant states, likely due to residual flux crosstalk and waveform distortion, as discussed in Fig.~\ref{fig:phase_vector_exp}. These effects also contribute to the discrepancy between experiment and simulation.

\subsection{Decay with alternating-phase waveguide coupling - Data and modeling}
\label{sec:SI_extradata-out-of-phase}

\begin{figure}[!htbp]
    \centering
    \includegraphics[width=0.7\columnwidth]{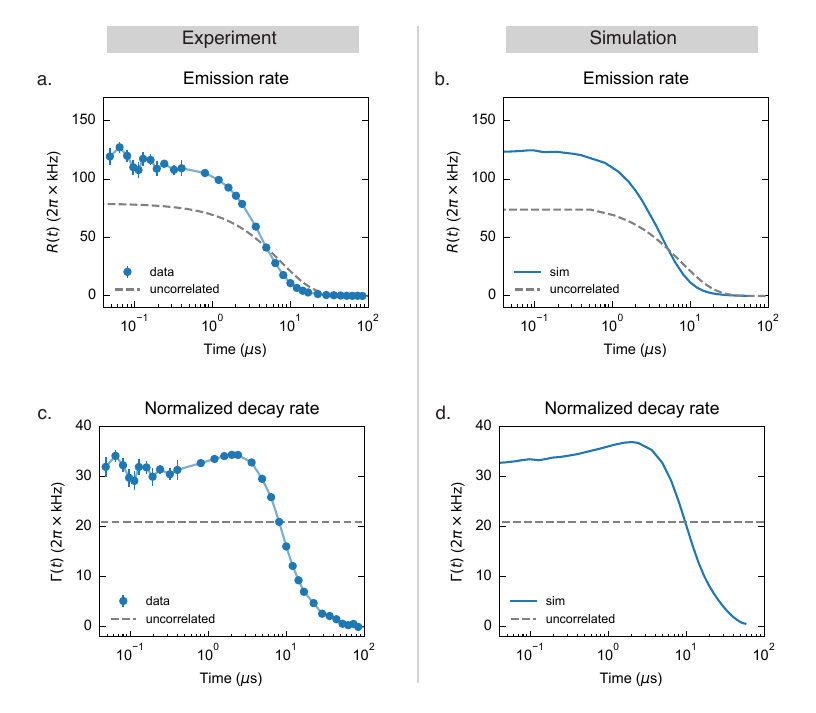}
    \caption[]{Emission and decay rates of the fully inverted state under collective decay with alternating-phase coupling ($\theta = \pi$).}
    \label{fig:outof_phase_exp_sim_combined} 
\end{figure}

In Fig.~\ref{fig:outof_phase_exp_sim_combined}, we present the experimental data and numerical simulations of the population dynamics for collective decay from the fully inverted state under alternating-phase coupling ($\theta = \pi$), corresponding to the experiment shown in Fig.~\ref{fig:5}(c-d). Compared to the $\theta=0$ case in Fig.~\ref{fig:4} and Fig.~\ref{fig:5}(a-b), the flux modulation frequency is detuned by $-0.5 \times 2\pi$~MHz ($\Delta_{\mathrm{eff}}^i \approx 2\pi\times\{65.8,\,20.1,\,-26.7,\,-77.2\}\,\mathrm{MHz}$) to further suppress unwanted local decay while maintaining the same engineered collective decay rates. The same detuning is used in both the experiment and simulation. We observe good agreement between the experimental data and numerical simulations.

\subsection{All pair-wise correlations - Data and modeling}
\label{sec:SI_extradata-correlation}

\begin{figure}[!htbp]
    \centering
    \makebox[\textwidth][c]{%
        \includegraphics[width=1.1\textwidth]{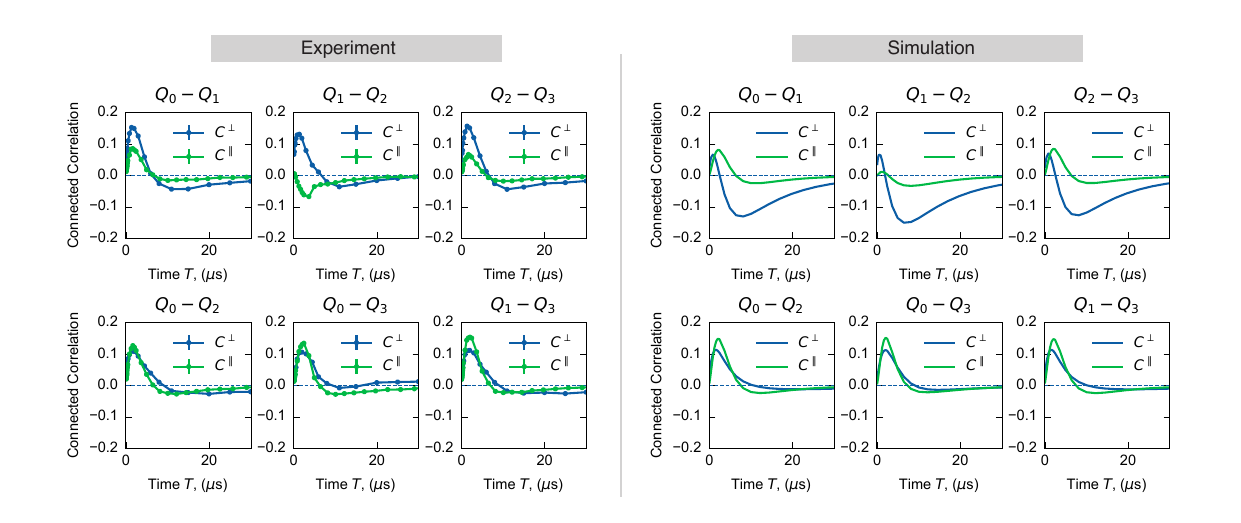}
    }
    \caption[]{Pair-wise correlations during uniform-phase collective decay ($\theta=0$).}
    \label{fig:in_phase_correlation} 
\end{figure}

\begin{figure}[!htbp]
    \centering
    \makebox[\textwidth][c]{%
        \includegraphics[width=1.09\textwidth]{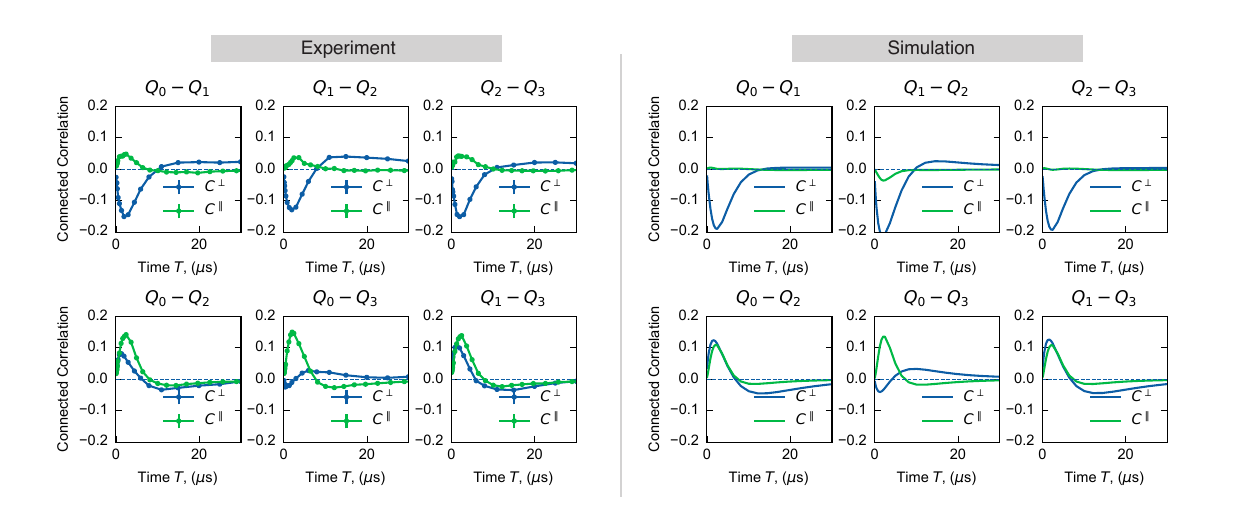}
    }
    \caption[]{Pair-wise correlations during alternating-phase collective decay ($\theta=\pi$).}
    \label{fig:outof_phase_correlation} 
\end{figure}

In Figs.~\ref{fig:in_phase_correlation} and \ref{fig:outof_phase_correlation}, we present the experimental data and numerical simulations of the pairwise two-qubit correlations (experiments in Fig.~\ref{fig:5}. In the experiment, we measure the expectation values of the single-site Pauli observables $X$, $Y$, and $Z$ for each qubit, as well as the two-body operators $XX$, $YY$, and $ZZ$ for each pair. We then calculate the connected symmetric transverse correlations $C^\perp_{ij} = (C^{xx}_{ij} + C^{yy}_{ij})/2$ and the connected longitudinal correlations $C^{\parallel}_{ij}= C^{zz}_{ij}$, where $C^{\alpha\beta}_{ij} = \langle \sigma^\alpha_i \sigma^\beta_j\rangle - \langle \sigma^\alpha_i\rangle\langle \sigma^\beta_j\rangle$.

We investigate possible systematic errors in the transverse correlation measurements arising from phase errors in the single-qubit $\pi/2$ rotations by comparing correlations measured in the $XX$, $YY$, $X(-X)$, and $Y(-Y)$ settings, where the minus sign is implemented using a virtual-$Z$ gate. Ideally, flipping the sign of one measurement axis reverses the sign of the corresponding correlator, such that $C^{X(-X)}_{ij}=-C^{XX}_{ij}$ and similarly for the $Y$ basis. We therefore estimate the systematic phase error from the residual quantity $C^{XX}_{ij}+C^{X(-X)}_{ij}$, whose nonzero offset is directly sensitive to relative phase errors between the sequentially applied single-qubit rotations. We additionally compare the measured $XX$ and $YY$ correlations and find their difference remains close to zero, consistent with theoretical expectations for decay from the fully inverted state. Experimentally, the resulting systematic error in the transverse correlations is below $0.005$, small compared to the typical correlation amplitudes observed during the superradiant and subradiant dynamics, while the statistical uncertainties are smaller than these systematic errors.
\smallskip

\textit{\textbf{Calculating many-body energy from pair-wise correlations:}}
In Fig.~\ref{fig:5}(a,c), we show the energy spectra and the measured $E(t)$ versus $N(t)$ trajectories.
To reconstruct the energy from measured correlations, we derive an effective Hamiltonian in the two-level qubit subspace that retains the effects of the finite transmon anharmonicity $U$. In particular, states with multiple excitations acquire energy shifts from virtual transitions to higher transmon levels. The resulting effective Hamiltonian is
\begin{equation}
    H/\hbar =
    \sum_i J_{i,i+1}
    \left(
    \sigma_i^+ \sigma^-_{i+1}
    + \sigma_i^- \sigma^+_{i+1}
    \right)
    +
    \sum_i J_{i,i+2}
    \left(
    \sigma_i^+ \sigma^-_{i+2}
    + \sigma_i^- \sigma^+_{i+2}
    \right)
    +
    \sum_{i,j}
    \chi_{i,j}
    \sigma_i^+\sigma_i^-
    \sigma_j^+\sigma_j^- ,
\end{equation}
or equivalently,
\begin{equation}
    H/\hbar =
    \sum_{i,j}
    \frac{J_{i,j}}{2}
    \left(
    \sigma_i^x \sigma_j^x
    + \sigma_i^y \sigma_j^y
    \right)
    +
    \sum_{i,j}
    \frac{\chi_{i,j}}{4}
    \left(
    1-\sigma_i^z
    \right)
    \left(
    1-\sigma_j^z
    \right),
\end{equation}
with
\begin{equation}
 \chi =  2\pi\times \begin{pmatrix}
     0.000 & 0.289 & 0.003 & 0.000\\
     0.289 & 0.000 & 0.289 & 0.003\\
     0.003 & 0.289 & 0.000 & 0.288\\
     0.000 & 0.003 & 0.288 & 0.000
 \end{pmatrix} \mathrm{MHz},
\end{equation}
calculated perturbatively. We verify that the energy spectrum of this effective Hamiltonian agrees well with that of the original multi-level transmon Hamiltonian.

This effective description allows reconstruction of the instantaneous many-body energy from the measured two-body correlations $\langle X_iX_j\rangle$, $\langle Y_iY_j\rangle$, $\langle Z_iZ_j\rangle$, and single-qubit expectation values $\langle Z_i\rangle$.


\input{supplementary.bbl}

%% file: supplementary.bbl
%